\documentclass[reprint, prd]{revtex4-2}
\usepackage{amsmath, amssymb}
\usepackage{graphicx,color,float,tabularx,siunitx}
\usepackage[caption=false]{subfig}
\usepackage{placeins}
\usepackage{rviewport}
\usepackage{orcidlink}
\definecolor{colorLink}{rgb}{0,0,180} 
\usepackage{hyperref}
\hypersetup{
   colorlinks = true,
   citecolor  = colorLink,
   urlcolor   = colorLink,
   linkcolor  = colorLink,
}

\DeclareSIUnit\year{yr}
\usepackage{subfig}
\usepackage{float} % for [H] specifier
\usepackage{booktabs}
\usepackage{multirow}
\usepackage{siunitx}

\usepackage[sort&compress]{natbib}

\usepackage[shortcuts]{extdash}

\begin{document}
\title{Observational Constraints on $f(R,T)$ gravity coupled with NED from Black Hole Quasi Periodic Oscillations}

\author{Bidyut Hazarika \orcidlink{0009-0007-8817-1945}$^1$}

\email{rs\_bidyuthazarika@dibru.ac.in}

\author{Abhishek Baruah \orcidlink{0009-0006-2069-0872}$^{1,2}$}
\email{rs\_abhishekbaruah@dibru.ac.in} 

\author{Prabwal Phukon \orcidlink{0000-0002-4465-7974}$^{1,3}$}
\email{prabwal@dibru.ac.in}

	\affiliation{$^1$Department of Physics, Dibrugarh University, Dibrugarh, Assam, 786004.\\$^2$Department of Physics, Patkai Christian College, Ch\"umukedima, Nagaland, 797103\\$^3$Theoretical Physics Division, Centre for Atmospheric Studies, Dibrugarh University, Dibrugarh, Assam, 786004.\\}

%\date{}
\begin{abstract}

In this work, we examine the influence of nonlinear electrodynamics (NED) and $f(R,T)$ gravity on quasi-periodic oscillations (QPOs) around a magnetically charged black hole. By analyzing the effective potential, specific energy, and angular momentum of circular orbits, we study how the NED parameter $\alpha$ and the gravitational coupling parameter $\beta$ affect orbital dynamics. Increasing $\alpha$ leads to systematic deviations from the Reissner–Nordstr\"om behavior, reflected in shifts in the ISCO radius and orbital frequencies. We further explore QPO-generating radii using the Relativistic Precession (RP), Warped Disk (WD), and Epicyclic Resonance (ER2–ER4) models. We further extend our analysis by performing an MCMC-based parameter estimation using QPO data from black holes spanning a wide range of mass scales. This approach yields consistent observational constraints on the model parameters.
\end{abstract}

%\pacs{}

\maketitle                                                                      

\section{Introduction}
Black holes stand among the most intriguing and profound predictions of General Relativity (GR), offering deep insights into the nature of spacetime and gravitation. Their existence not only highlights the theoretical richness of GR but also challenges our understanding of fundamental physics under extreme conditions. Since its formulation by Einstein in 1915, General Relativity (GR) has served as a rigorous and comprehensive framework for describing the curvature of spacetime. A landmark validation of GR was achieved through the detection of gravitational waves originating from binary black hole mergers by the LIGO collaboration, representing a pivotal breakthrough in the field of gravitational physics \cite{ligo}.
 This milestone was soon followed by the ground breaking observations made by the Event Horizon Telescope (EHT), which produced the first-ever images of super-massive black holes initially in the galaxy M87*, and subsequently in Sagittarius A* (SgrA*), located at the center of the Milky Way \cite{m87a, m87b, m87c, m87d, m87e, m87f}. These unprecedented visualizations unveiled a dark central silhouette encircled by a bright photon ring, whose morphology encodes vital information about the black hole's properties and the underlying gravitational theory \cite{Shadow1,Shadow2,Shadow3,Shadow4}.
 A substantial body of research suggests that the geometric features of black hole shadows, along with the properties of photon spheres, provide powerful diagnostic tools for probing and constraining potential departures from General Relativity. As such, they serve as critical observables for evaluating the viability of alternative gravitational theories \cite{Shadow5,Shadow6,Shadow7,Shadow8,Shadow9,Shadow10,Shadow11,Shadow12}.

 Magnetic fields are prevalent in astrophysical settings and exert a profound influence on the behavior and dynamics of charged matter in the vicinity of compact astrophysical objects.
 Their impact becomes especially pronounced near black holes, where the interplay of intense gravitational and electromagnetic fields can markedly alter the trajectories of test particles.
 Significantly, the coupling between a black hole’s external magnetic field and the dipole moment of a particle offers valuable insights into the dynamical behaviour of matter in the vicinity of magnetized compact objects.

Foundational studies, such as Wald’s \cite{1} analytical solution describing electromagnetic fields in the vicinity of a Kerr black hole embedded in a uniform magnetic background, established the groundwork for extensive explorations of magnetized black hole geometries. Later investigations broadened this framework to encompass a diverse range of magnetic field configurations which includes dipolar and split-monopole structures and explored their effects across various spacetime backgrounds, analyzing the dynamics of both neutral and charged particles subjected to these fields. Such inquiries are especially critical for elucidating the physics of accretion processes, mechanisms of particle acceleration, and the formation of relativistic jets.

Analysing particle dynamics in the vicinity of black holes is fundamental to probing their physical properties and spacetime geometry. Over the years, a considerable body of work has focused on the trajectories of both massive and massless particles within a wide range of parametrized black hole metrics \cite{2,3,4,5,6,7,8,9,10,11,12}.
  
 Orbital and epicyclic frequencies within axisymmetric and stationary spacetimes have been the focus of extensive investigation, owing to their significance in characterizing particle motion in black hole surroundings \cite{11}. Foundational contributions offered exact analytical formulations of geodesics, thereby establishing a basis for more sophisticated and comprehensive analyses \cite{13}.
 Subsequent research expanded upon these results by incorporating the motion of charged test particles within spacetimes subject to both electric and magnetic field influences \cite{14,15}. Remarkably, recent studies have shown that the interplay between electric charge and external magnetic fields in Reissner–Nordström geometry can effectively emulate the dynamics characteristic of black holes possessing intrinsic magnetic charge \cite{16}, thereby introducing additional layers of complexity to particle motion in such environments.
 
Quasi-periodic oscillations (QPOs), detected in the X-ray emissions of black holes and neutron stars, have emerged as powerful diagnostics for probing the physics of strong-field gravity. Characterized by nearly periodic fluctuations in luminosity, these oscillations are thought to originate from fundamental mechanisms such as accretion disk dynamics and relativistic gravitational phenomena. In particular, the observation of twin-peak QPOs in specific astrophysical systems has spurred significant research efforts to elucidate their underlying origin, frequently attributed to resonant interactions or oscillatory modes within the accretion disk. This has underscored the pressing need for more sophisticated theoretical frameworks, bolstered by high-precision observational data, to accurately capture the complex dynamics governing these phenomena.
 Since their initial discovery through spectral and timing analyses in X-ray binary systems \cite{17}, QPOs have been extensively investigated from both observational and theoretical perspectives. Among the array of proposed models, those grounded in the dynamics of particles in curved spacetime have garnered significant attention. In such frameworks, the quasi-periodic nature of the oscillations is attributed to perturbations in the trajectories of charged test particles, which in turn influence the morphology and temporal evolution of the accretion flow \cite{18,19,20,21,22,23,24,25,26,27,28,29,30,31,32}.
Recent computational investigations have explored the underlying mechanisms driving QPO formation in black hole spacetimes by solving the general relativistic hydrodynamic equations \cite{33}, particularly in backgrounds such as Kerr and hairy black holes. These simulations demonstrate that perturbations in the accreting plasma can give rise to spiral shock structures, which exhibit a strong correlation with the emergence of QPO phenomena \cite{34,35,36}.
 Likewise, theoretical models employing Bondi–Hoyle–Lyttleton accretion have demonstrated that shock cones generated within intense gravitational fields can give rise to distinct QPO signatures \cite{37,38,39,40,41}. These approaches have proven successful in accounting for observed QPO behavior in sources such as GRS 1915+105 \cite{42}, and further provide predictive insights into potential QPO characteristics in the vicinity of supermassive black holes, including M87* \cite{43}. The dynamics of test particles and the resulting quasi-periodic oscillations (QPOs) in the vicinity of black holes have been thoroughly examined across a wide range of studies; see, for example, Refs.~~\cite{c1,c2,c3,c4,c6,c6,c7} for a representative selection. In particular, Ref.~~\cite{c2} investigates black hole solutions sourced by nonlinear electrodynamics, analyzing their QPO signatures with a focus on regular rotating black holes, thereby highlighting the interplay between modified electrodynamics and observable oscillatory phenomena.

Among the numerous attempts to address the limitations of General Relativity (GR) in explaining phenomena such as dark matter, cosmic inflation, and the late-time accelerated expansion of the universe, modified gravity theories have garnered significant attention. One such extension is $f(R,T)$ gravity, where the Einstein-Hilbert action is generalized to include an arbitrary function of both the Ricci scalar $R$ and the trace $T$ of the energy-momentum tensor 
$T_{\mu \nu}$ \cite{frt1,frt2}. The inclusion of the 
$T$-dependence introduces explicit curvature-matter couplings, leading to a non-conserved energy-momentum tensor and the emergence of an additional force \cite{frt3,frt4}. This deviation from geodesic motion enables $f(R,T)$ gravity to model a variety of physical phenomena that remain unexplained within the standard GR framework. Consequently, this theory has been employed in a wide array of contexts, including black hole thermodynamics \cite{frt5}, energy condition analyses \cite{frt6,frt7}, compact star configurations \cite{frt8,frt9,frt10,frt11}, gravastar models \cite{frt12,frt13}, cosmological evolution \cite{frt14,frt15,frt16,frt17}, and wormhole geometries \cite{frt18,frt19,frt20,frt21,frt22}.

In parallel, nonlinear electrodynamics (NLED) has emerged as a compelling matter source for constructing singularity-free black hole solutions. Unlike classical Maxwell theory, NLED incorporates higher-order field corrections that become prominent in strong-field regimes, such as those near compact astrophysical objects. A seminal advancement in this area was the reinterpretation of the Bardeen black hole as a regular solution of GR sourced by NLED, as demonstrated by Ayón-Beato and García \cite{nled1,nled2}. These models circumvent the central singularity by modifying the small-scale structure of spacetime through electromagnetic self-interactions. Numerous subsequent studies have explored a range of NLED-inspired regular black holes within GR \cite{nled3,nled4,nled5,nled6,nled7}. Beyond their role in resolving singularities, NLED frameworks have also been investigated in cosmological settings, where they have been shown to drive accelerated expansion without invoking dark energy \cite{nled8,nled9}. Thus, NLED serves not only as a physically motivated modification to electrodynamics but also as a robust tool in addressing key challenges in gravitational and cosmological physics.

In the following paragraph, we present a concise summary of the nonlinear electrodynamics (NED) black hole solution within the $f(R,T)$ gravity framework adopted in this study.
\\
The action function characterizing the nonlinear electrodynamics (NED) black hole within the framework of $f(R,T)$ gravity, as employed in this study, is formulated in accordance with the construction outlined in ~\cite{nled10}:  
%\begin{equation}\label{eq2}
%S = \int d^4x \sqrt{-g} \left[ f(R,T)+2\kappa^2 \mathcal{L}_{NLED}(F) \right],
%\end{equation}  
%\textcolor{red}{The quantity $R$ denotes the Ricci scalar, while $T$ corresponds to the trace of the energy-momentum tensor. The term $\mathcal{L}_{\text{NLED}}(F)$ represents the Lagrangian density for the nonlinear electrodynamics (NLED), which is a function of the electromagnetic invariant $F = \frac{1}{4} F_{\mu\nu}F^{\mu\nu}$. Here, $F_{\mu\nu} = \partial_\mu A_\nu - \partial_\nu A_\mu$ defines the antisymmetric Maxwell–Faraday field strength tensor, and $A_\mu$ denotes the electromagnetic four-potential.}
%In this work, we focus on static, spherically symmetric black hole solutions arising from the coupling of nonlinear electrodynamics (NED) to modified \( f(R,T) \) gravity. The starting point is the action functional proposed in Ref.~\cite{nedfrt1}, given by:
\begin{equation}
S = \int d^4x \sqrt{-g} \left[ f(R,T) + 2\kappa^2 \mathcal{L}_{\text{NLED}}(F) \right],
\label{eq:action}
\end{equation}
where \( g \) is the determinant of the metric tensor \( g_{\mu\nu} \), \( R \) is the Ricci scalar, and $T$ is the trace of the energy-momentum tensor. We set \( \kappa^2 = 8\pi \). The Lagrangian density \( \mathcal{L}_{\text{NLED}}(F) \) encodes the nonlinear electromagnetic interaction and depends on the field invariant
\begin{equation}
F = \frac{1}{4} F^{\mu\nu} F_{\mu\nu},
\end{equation}
where \( F_{\mu\nu} = \partial_\mu A_\nu - \partial_\nu A_\mu \) is the antisymmetric field strength tensor derived from the electromagnetic potential \( A_\mu \).
We adopt the specific form of the gravitational Lagrangian:
\begin{equation}
f(R,T) = R + \beta T,
\label{eq:frt}
\end{equation}
where \( \beta \) is a constant coupling parameter quantifying the nonminimal interaction between curvature and matter.
 For the electromagnetic sector, the equation of motion follow from variation with respect to \( A_\mu \), leading to a modified Maxwell equation:
\begin{equation}
\nabla_\mu \left[ \left( 2 f_T(R,T) \mathcal{L}_{\text{FF}} F - \kappa^2 \mathcal{L}_F \right) F^{\mu\alpha} \right] = 0,
\label{eq:maxwell}
\end{equation}
where \( \mathcal{L}_F \equiv \partial \mathcal{L}_{\text{NLED}}(F)/\partial F \), and \( f_T = \partial f(R,T)/\partial T \).
Variation of the action \eqref{eq:action} with respect to the metric tensor yields the modified Einstein field equations:
\begin{equation}
\begin{split}
&f_R R_{\mu\nu} - \frac{1}{2} f g_{\mu\nu} + (g_{\mu\nu} \Box - \nabla_\mu \nabla_\nu) f_R\\
& = T_{\mu\nu} - f_T \left( T_{\mu\nu} + \Theta_{\mu\nu} \right),
\end{split}
\label{eq:field}
\end{equation}
where \( f_R \equiv \partial f(R,T)/\partial R \), \( \Box = g^{\mu\nu} \nabla_\mu \nabla_\nu \), $T_{\mu \nu}=-\frac{2}{\sqrt{-g}}\frac{\delta \mathcal{L}_{mat}\sqrt{-g}}{\delta g^{\mu \nu}}$ where $\mathcal{L}_{mat}$ is the density of matter Lagrangian and \( \Theta_{\mu\nu} \equiv g^{\alpha\beta} \delta T_{\alpha\beta}/\delta g^{\mu\nu} \). For the NED source, the energy-momentum tensor is:
\begin{equation}
T_{\mu\nu} = g_{\mu\nu} \mathcal{L}_{\text{NLED}}(F) - \mathcal{L}_F F_{\mu\rho} F_{\nu}{}^{\rho},
\label{eq:tmunu}
\end{equation}
and \begin{equation}
\begin{split}
&\Theta_{\mu\nu} = - g_{\mu\nu} \mathcal{L}_{\text{NLED}}(F)\\
& + F_{\mu\rho} F_{\nu}^{\rho} \left( \mathcal{L}_F(F) - \frac{1}{2} F_{\alpha \beta}F^{\alpha \beta} \mathcal{L}_{FF}(F) \right),
\end{split}
\end{equation}
with
\begin{equation}  \mathcal{L}_{F} = \frac{\partial \mathcal{L}_{\text{NLED}}}{\partial r}\left( \frac{\partial F}{\partial r}\right)^{-1}, \quad \mathcal{L}_{FF} = \frac{\partial \mathcal{L}_{\text{F}}}{\partial r}\left( \frac{\partial F}{\partial r}\right)^{-1} 
\end{equation}
To solve these equations, we consider the standard static and spherically symmetric line element:
\begin{equation}
ds^2 = A(r)\, dt^2 - \frac{dr^2}{A(r)} - r^2 \left( d\theta^2 + \sin^2\theta\, d\phi^2 \right),
\end{equation}
and a purely magnetic field configuration, where the only non-vanishing component of \( F_{\mu\nu} \) is:
\begin{equation}
F_{23} =-F_{32}= q \sin \theta 
\end{equation}
where $F$ as electromagnetic scalar is given as $F=q^2/2r^4$\\
We now introduce a power-law NED model:
\begin{equation}
\mathcal{L}_{\text{NLED}}(F) = f_0 + F + \alpha F^p,
\label{eq:nled}
\end{equation}
where \( f_0 \) is an integration constant that contributes to an effective cosmological term, \( \alpha \) controls the nonlinearity strength, and \( p \in \mathbb{R} \) is the power index.
Substituting this form into the field equations and assuming \( A(r) = B(r) \), the solution for the metric function becomes:
\begin{equation}
\begin{split}
&A(r) = 1 - \frac{2M}{r} + \frac{q^2}{r^2} - \frac{\Lambda_{\text{eff}}}{3} r^2 \\
&+ \frac{2^{1-p}}{3 - 4p} \alpha \left[2\beta(p - 1) - 1\right] q^{2p} r^{2 - 4p}, 
\end{split}
\label{eq:metricA}
\end{equation}
\begin{equation}
\Lambda_{\text{eff}} \equiv 2(2\beta + 1) f_0. 
\label{eq:lambdaeff}
\end{equation}
For the illustrative case \( p = 2 \), Eq.~\eqref{eq:metricA} simplifies to:
\begin{equation}
A(r) = 1 - \frac{2M}{r} + \frac{q^2}{r^2} - \frac{\Lambda_{\text{eff}}}{3} r^2 - \frac{\alpha(2\beta - 1) q^4}{10 r^6}.
\label{eq:metricP2}
\end{equation}
The corresponding Lagrangian in terms of the radial coordinate becomes:
\begin{align}
\mathcal{L}_{\text{NLED}}(r) &=f_0+\frac{q^2}{2r^4}+\alpha 2^{-p}\left(\frac{q^2}{r^4}\right)^p= f_0 + \frac{q^2}{2r^4} + \frac{\alpha q^4}{4r^8} \\
\mathcal{L}_F(r) &=1+\alpha 2^{1-p}p\left(\frac{q^2}{r^4}\right)^{p-1}= 1 + \alpha    \frac{q^2}{r^4}, \\
\mathcal{L}_{FF}(r) &=\alpha 2^{2-p}(p-1)p\left(\frac{q^2}{r^4}\right)^{p-2}= 2\alpha
\end{align}
Rewriting the Lagrangian as a function of \( F \) using \( F = \frac{q^2}{2r^4} \), we obtain:
\begin{equation}
\mathcal{L}_{\text{NLED}}(F) = \frac{\Lambda_{\text{eff}}}{4\beta + 2} + F + \alpha F^2.
\end{equation}
In the weak-field limit \( F \ll 1 \), the Lagrangian reduces to:
\begin{equation}
\mathcal{L}_{\text{NLED}}(F) \approx \frac{\Lambda_{\text{eff}}}{4\beta + 2} + F,
\end{equation}
recovering the linear Maxwell form with a small effective vacuum energy. In the strong-field regime, nonlinear terms become dominant:
\begin{equation}
\mathcal{L}_{\text{NLED}}(F) \approx \frac{\Lambda_{\text{eff}}}{4\beta + 2} + F + \alpha F^2.
\end{equation}
This black hole metric provides a construction that generalizes the Reissner–Nordstr\"om–AdS geometry, incorporating both nonlinear electromagnetic corrections and matter-curvature coupling effects from \( f(R,T) \) gravity. Setting \( \alpha = 0 \) and \( \beta = 1/2 \) recovers the standard RN–AdS solution. The analysis shows that the parameters \( \alpha \) and \( \beta \) play key roles in shaping the spacetime structure and may encode observable imprints, such as those found in quasi-periodic oscillation (QPO) frequencies.

The primary objective of this study is to examine whether the distinctive features exhibited by the spacetime geometry in the extremal regimes of the nonlinear electrodynamics (NED) parameter $\alpha$ also manifest in the behavior of quasi-periodic oscillations (QPOs). In particular, we aim to determine whether analogous trends emerge in the QPO frequency spectra as $\alpha$ varies , and how the NED parameter, in conjunction with the matter-curvature coupling constant $\beta$ from the $f(R,T)$ gravity framework, modulates these features. Our principal goal is to discern the characteristic ``imprints" of nonlinear electrodynamics embedded in the QPO signatures by systematically exploring these dynamical behaviors.
\\
In this work, we investigate the dynamics of neutral test particles in the background of a static, spherically symmetric magnetically charged black hole solution derived from nonlinear electrodynamics (NED) coupled to $f(R,T)$ gravity. The central objective is to uncover the signatures of the NED nonlinearity parameter $\alpha$ and the gravitational coupling parameter $\beta$ on the quasi-periodic oscillations (QPOs) associated with such black holes. To achieve this, we systematically examine the orbital characteristics of test particles, specifically the effective potential, specific angular momentum, and energy of stable circular orbits and assess how these quantities evolve with variations in $\alpha$ and $\beta$. Our analysis reveals a smooth deformation of the spacetime geometry away from the Reissner–Nordstr\"om (RN) limit as $\alpha$ increases, manifested through pronounced shifts in the location of the innermost stable circular orbit (ISCO) and modifications to the Keplerian frequency profile.
\\
To further elucidate the impact of these parameters, we explore the behavior of QPO-generating radii across several phenomenological models, including the Relativistic Precession (RP), Warped Disk (WD), and Epicyclic Resonance (ER) frameworks. We complement this theoretical investigation with a Markov Chain Monte Carlo (MCMC) analysis, utilizing observational QPO data from a wide range of black hole systems encompassing stellar-mass, intermediate-mass, and supermassive categories. The resulting posterior distributions yield consistent and robust constraints on $\alpha$ and $\beta$, indicating that both nonlinear electrodynamic effects and matter-curvature interactions in modified gravity imprint discernible features on the QPO structure of magnetically charged black holes across all mass scales. These findings provide compelling evidence for the viability of NED-coupled $f(R,T)$ gravity as an alternative avenue for probing strong-field astrophysical phenomena.

\section{Particle Dynamics around NED Black Holes in $f(R,T)$ Gravity} 
\label{null}
\subsection{Equations of Motion}
In this section, we investigate the trajectories of electrically neutral test particles in the spacetime surrounding a charged black hole sourced by nonlinear electrodynamics (NED) in $f(R,T)$ gravity. The evolution of these particles is determined by the following Lagrangian formalism given as:

\begin{equation}
    L_p = \frac{1}{2} m g_{\mu\nu} \dot{x}^\mu \dot{x}^\nu,
\end{equation}
Here, $m$ represents the rest mass of the particle, and the dot over it signifies differentiation with respect to the proper time $\tau$. It is important to emphasize that $x^\mu(\tau)$ describes the particle's worldline, parameterized by its proper time $\tau$, whereas the corresponding four-velocity is given by $u^\mu = \frac{dx^\mu}{d\tau}$.

In a spherically symmetric spacetime, the presence of two Killing vector fields $\xi^\mu = (1, 0, 0, 0)$ and $\eta^\mu = (0, 0, 0, 1)$ reflects the invariance of the geometry under time translations and axial rotations, respectively. As a consequence, the motion of a test particle admits two conserved quantities: the total energy $E$ and the angular momentum $L$, which are expressed as follows:
\begin{equation}
\mathcal{E}=-g_{tt}  ~\Dot{t},\quad
\mathcal{L}=g_{\phi \phi}~ \Dot{\phi}.
\label{energy}
\end{equation}
In Eq. (\ref{energy}), the quantities $\mathcal{E}$ and $\mathcal{L}$ denote the specific energy and specific angular momentum of the test particle, i.e., energy and angular momentum per unit mass. The corresponding equation of motion can be obtained by invoking the normalization condition on the four-velocity and is given as:

\begin{align}\label{normal}
    g_{\mu\nu} u^\mu u^\nu = \delta,
\end{align}
Here, $\delta = 0$ and $\delta = \pm 1$ distinguish between geodesic motion for massless and massive particles, respectively. In particular, $\delta = +1$ pertains to spacelike geodesics, while $\delta = -1$ characterizes timelike geodesics. For massive particles, the dynamics are dictated by timelike geodesics of the underlying spacetime geometry, and the associated equations of motion can be derived by utilizing Eq.~(\ref{normal}).

By employing Eqs.~(\ref{energy}) and (\ref{normal}), the equation governing the particle’s motion confined to a constant plane can be reformulated in the following manner given below:
\begin{align}
    \dot{r}^2=\mathcal{E}+g_{tt}\Big(1+\frac{{\mathcal{L}^2}}{r^2 }\Big),
\end{align}
In a static, spherically symmetric spacetime, a particle that initiates its motion in the equatorial plane will remain confined to that plane for the entirety of its trajectory due to the symmetry of the system. By imposing this restriction by specifically, setting $\theta = \frac{\pi}{2}$ and $\dot{\theta} = 0$, the radial component of the equation of motion simplifies and can be expressed as follows:
\begin{align}
    \dot{r}^2 = \mathcal{E}^2 - V_\text{eff},
\end{align}
By imposing the standard criteria for circular motion, namely $\dot{r} = 0$ and $\ddot{r} = 0$, one can obtain the following set of equations, which characterize stable circular trajectories in the given spacetime background, given as:

\begin{equation}
    \dot{r}=0, \quad V_\text{eff}=\mathcal{E}^2
     \label{cosnt}
\end{equation}\\

\begin{figure}[!h]
\centering
\includegraphics[width=0.95\linewidth]{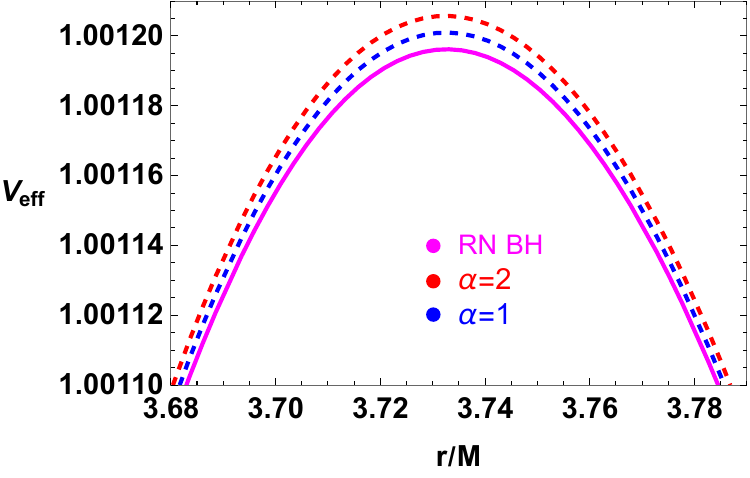}
\vspace{-0.3cm}
\caption{Radial profile of the effective potential for varying nonlinearity parameter $\alpha$, with fixed $Q = 0.5$, $\beta = 0.002$, and power-law index $p = 2$.} 
\label{fig2}
\end{figure}
\begin{figure}[!h]
\centering
\includegraphics[width=0.95\linewidth]{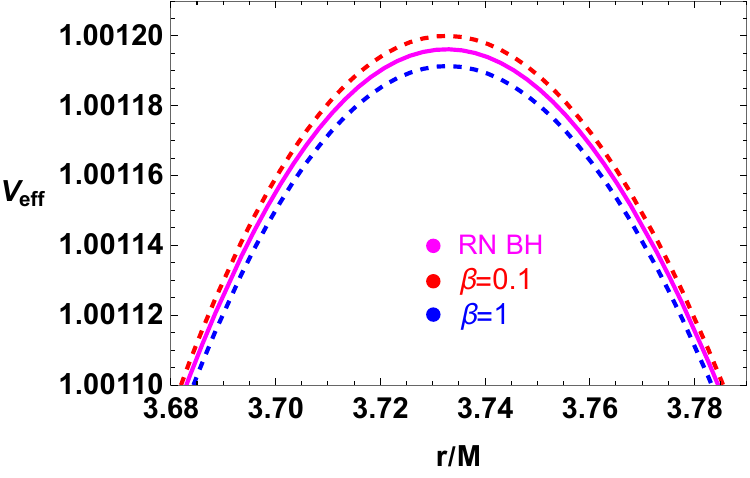}
\vspace{-0.3cm}
\caption{Radial profile of the effective potential for varying coupling parameter $\beta$, with fixed $\alpha = 1$, $Q = 0.5$, and power-law index $p = 2$.} 
\label{fig3}
\end{figure}
\vspace{0.5cm}
The effective potential that governs the radial dynamics of a test particle constrained to motion in the equatorial plane is expressed as follows:

\begin{align}
    V_\text{eff} = f(r) \left(1 + \frac{\mathcal{L}^2}{r^2}\right).
\end{align}
\begin{figure*}[!t]
\centering
\includegraphics[width=0.32\linewidth]{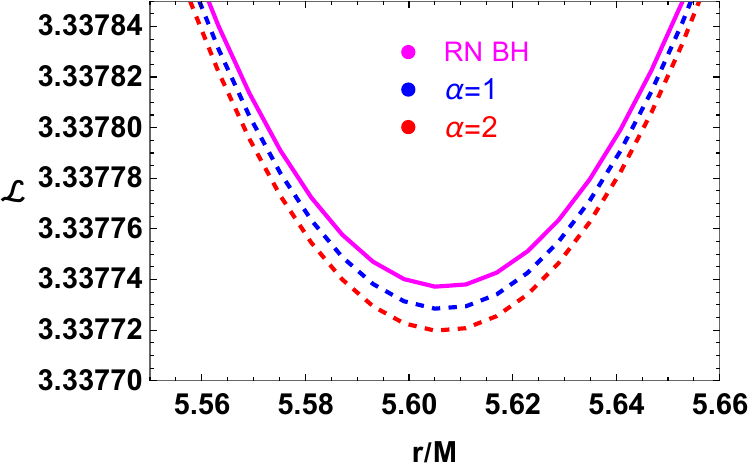}  % Adjust width to fit three figures
\includegraphics[width=0.32\linewidth]{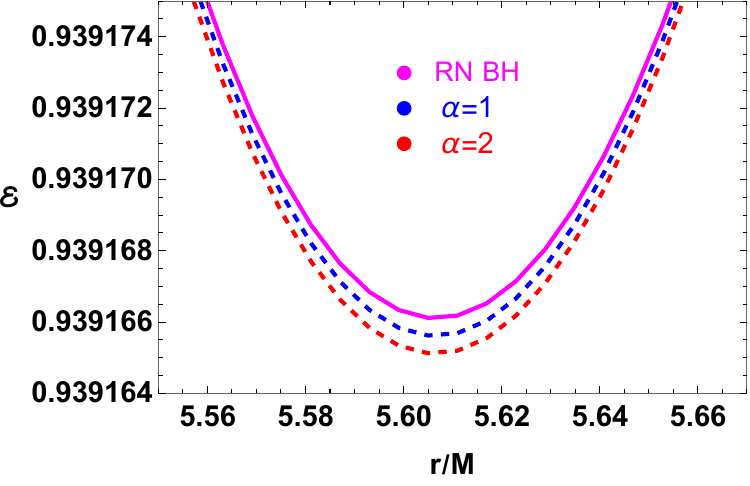}
\vspace{-0.3cm}
\caption{Radial variation of the specific angular momentum and specific energy for circular orbits, shown for different values of the NED parameter $\alpha$, with the magnetic charge fixed at $Q = 0.5$.}
\label{fig4}
\end{figure*}
\begin{figure*}[!t]
\centering
\includegraphics[width=0.32\linewidth]{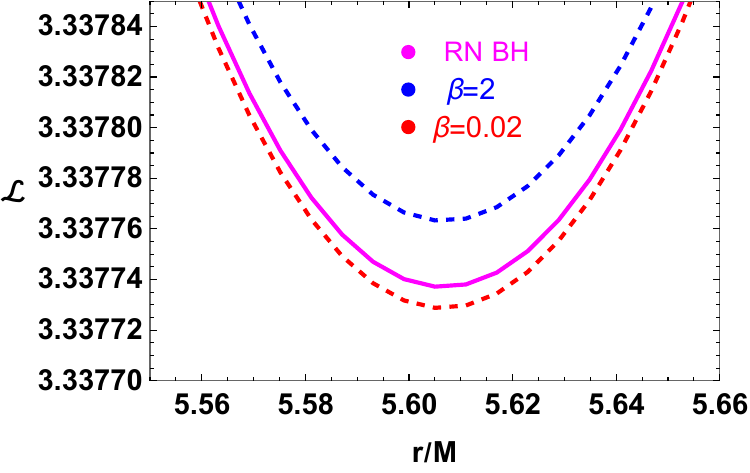}  % Adjust width to fit three figures
\includegraphics[width=0.32\linewidth]{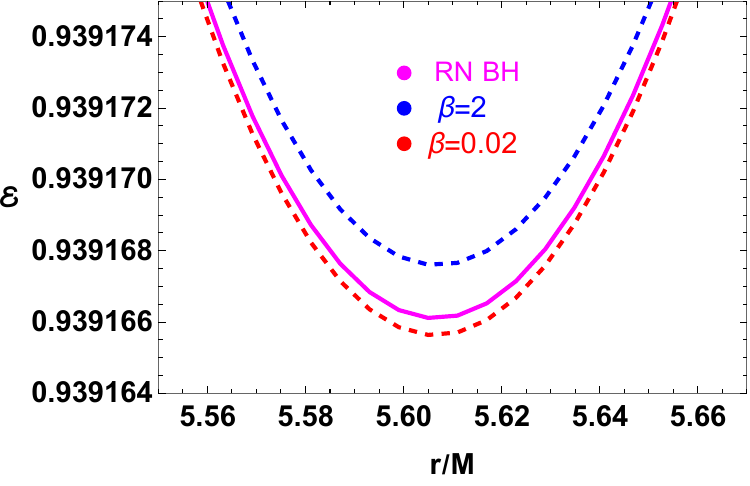}
\vspace{-0.3cm}
\caption{Radial variation of the specific angular momentum and specific energy for circular orbits, shown for different values of the coupling parameter $\beta$, with the magnetic charge fixed at $Q = 0.5$.
}
\label{fig5}
\end{figure*}
Figure \ref{fig2} illustrates the radial behavior of the effective potential for electrically neutral test particles in a magnetically charged black hole spacetime described by nonlinear electrodynamics (NED) in $f(R,T)$. The plot compares the effective potential for different values of the non-linearity parameter $\alpha$ (specifically, $\alpha = 1$ and $\alpha = 2$) with the standard Reissner–Nordstr\"om (RN) black hole case.
Figure \ref{fig3} shows the corresponding variation in the effective potential for different values of the matter-curvature coupling parameter $\beta$ (with $\beta = 0.1$ and $\beta = 1$), again contrasted with the RN black hole.
In both figures, the magnetic charge is held fixed at $Q = 0.5$, and the power-law index in the NED Lagrangian is set to $p = 2$. As seen from the plots, increasing either $\alpha$ or $\beta$ reduces the peak height of the effective potential relative to the RN case. This behavior suggests that stronger non-linearity in the electromagnetic sector or a more prominent matter-curvature coupling leads to a softening of the potential barrier.
The extrema of the effective potential which is its minima and maxima corresponds to stable and unstable circular orbits, respectively. These features are directly influenced by the parameters $\alpha$ and $\beta$, which modify the underlying spacetime geometry and thereby the dynamics of neutral test particles.

Next, by employing the relations given in Eq.~(\ref{cosnt}), we obtain the expressions for the specific angular momentum and specific energy associated with circular orbits, which can be formulated as follows:

\begin{align}
    \mathcal{L}=\frac{-10 M r^7+\alpha  (3-6 \beta ) q^4 r^2+10 q^2 r^6}{10 r^5 (3 M-r)+4 \alpha  (2 \beta -1) q^4-20 q^2 r^4},
\end{align}
\begin{align}
    \mathcal{E}=\frac{\left(10 r^5 (r-2 M)+q^4 (\alpha -2 \alpha  \beta )+10 q^2 r^4\right)^2}{20 r^6 \left(5 r^5 (r-3 M)+\alpha  (2-4 \beta ) q^4+10 q^2 r^4\right)}
\end{align}
Figure \ref{fig4} presents the radial dependence of specific angular momentum $\mathcal{L}$ and energy $\mathcal{E}$ for circular orbits, illustrating the effect of the nonlinear electrodynamics (NED) parameter $\alpha$ in the context of $f(R,T)$ gravity.
The left panel illustrates the variation of the specific angular momentum $\mathcal{L}$ as a function of the radial coordinate $r/M$. The solid magenta line represents the Reissner–Nordström (RN) black hole, while the blue and red dashed curves show the effects of the nonlinear electrodynamics parameter $\alpha$ for $\alpha = 1$ and $\alpha = 2$, respectively.
 An increase in the value of $\alpha$ leads to a reduction in the minimum specific angular momentum relative to that of the Reissner–Nordstr\"om black hole. The middle panel depicts the radial variation of the specific energy $\mathcal{E}$ for circular orbits as a function of $r/M$, following the same color scheme as the left panel. The energy curve displays a minimum, signifying the location of the most energetically bound orbit.
Compared to the RN black hole, the nonlinear electrodynamics parameter $\alpha$ modifies the energy required for stable circular orbits, primarily affecting the depth of the energy minimum, while the position of the minimum remains nearly unchanged which mirrors the trend seen in the 
$\mathcal{L}$ vs. $r/M$ profile. \\
Figure \ref{fig5} presents the radial dependence of the specific angular momentum $\mathcal{L}$ and specific energy $\mathcal{E}$ for circular orbits, highlighting the influence of the matter-curvature coupling parameter $\beta$ in the context of $f(R,T)$ gravity.
The left panel illustrates how $\mathcal{L}$ varies with the radial coordinate $r/M$. The solid magenta curve corresponds to the Reissner–Nordström (RN) black hole, while the red and blue dashed curves represent the effects of the coupling parameter $\beta$ for $\beta = 2$ and $\beta = 0.02$, respectively. Increasing $\beta$ leads to a noticeable reduction in the minimum value of the specific angular momentum compared to the RN case.
The right panel shows the radial behavior of the specific energy $\mathcal{E}$ for circular orbits, following the same color coding. The energy profile exhibits a minimum, indicating the most energetically bound orbit. Similar to the angular momentum behavior, increasing $\beta$ lowers the depth of the energy minimum, while its radial location remains nearly unchanged reflecting the consistent trend observed in the $\mathcal{L}$ vs. $r/M$ plot.
 
\subsection{Innermost stable circular orbits (ISCO)}\label{secIII}

Solving the condition $V_{\text{eff}} = 0$ with respect to the radial coordinate $r$ enables the identification of points where the effective potential attains extremal values. Stable circular orbits are associated with local minima of the potential, characterized by $\partial_r^2 V_{\text{eff}}(r) > 0$, while instability arises when $\partial_r^2 V_{\text{eff}}(r) < 0$, indicating the presence of a local maximum.
The innermost stable circular orbit (ISCO) is determined by the condition $\partial_r^2 V_{\text{eff}}(r_{\text{ISCO}}) = 0$. However, in the case of black holes governed by nonlinear electrodynamics (NED) within the framework of $f(R,T)$ gravity, obtaining analytical solutions is nontrivial due to the presence of intricate hyperbolic functions in the metric. Therefore, we employ numerical methods to solve the relevant equations and present the resulting ISCO radius as a function of the coupling parameter $\beta$.
\\

\begin{figure}[!h]
\centering
\includegraphics[width=0.95\linewidth]{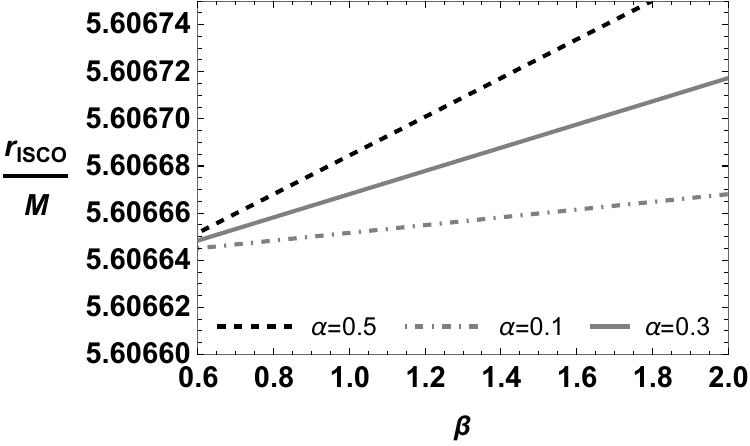}
\vspace{-0.3cm}
\caption{ISCO radius as a function of coupling parameter $\beta$ for NED black holes in $f(R,T)$ gravity.
 } 
\label{fig6}
\end{figure}
Figure \ref{fig6} illustrates the dependence of the ISCO radius on the coupling parameter $\beta$ for various values of the nonlinearity parameter $\alpha$ in the context of a black hole described by nonlinear electrodynamics in $f(R,T)$ gravity. The plot clearly shows that for all considered values of $\alpha$, the ISCO radius increases monotonically with increasing $\beta$, indicating a consistent trend in how matter-curvature coupling influences the orbital stability region.
 The observed trend indicates that the inclusion of nonlinear electrodynamics (NED) effects within $f(R,T)$ gravity leads to an outward shift of the ISCO radius. Moreover, for a fixed value of the coupling parameter $\beta$, the ISCO radius increases monotonically with increasing $\alpha$. This behavior reflects the influence of enhanced nonlinearity in the electromagnetic sector, which modifies the spacetime geometry and extends the region of orbital stability farther from the black hole.

\section{Fundamental frequencies}
In this section, we compute the fundamental frequencies that govern the motion of a test particle in the vicinity of a black hole sourced by nonlinear electrodynamics (NED) within the framework of $f(R,T)$ gravity. Specifically, we focus on the Keplerian orbital frequency, as well as the radial and vertical epicyclic frequencies associated with small perturbations around circular orbits.
 
\subsection{Keplerian frequencies}
The angular velocity of a test particle in circular motion around a black hole, as measured by an observer at spatial infinity, is referred to as the orbital or Keplerian frequency, denoted by $\Omega_\phi$. It is defined through the relation $\Omega_\phi = \frac{d\phi}{dt}$. Employing this definition, the general form of the orbital frequency in a static and spherically symmetric spacetime can be derived \cite{50}, and is expressed as follows:

\begin{align}
    \Omega_\phi = \sqrt{\frac{-\partial_r g_{tt}}{\partial_r g_{\phi\phi}}} = \sqrt{\frac{f'(r)}{2r}}.
\end{align}
For black holes governed by nonlinear electrodynamics (NED) within the framework of $f(R,T)$ gravity, the orbital frequency expression is modified to incorporate the effects of the nonlinearity parameter $\alpha$ and the matter-curvature coupling parameter $\beta$, and is given by:

\begin{align}
\Omega_\phi = \sqrt{\frac{M}{r^3}+\frac{3 \alpha  (2 \beta -1) Q^4}{10 r^8}-\frac{q^2}{r^4}}
\end{align}
By setting $Q = 0$, the angular velocity reduces to that of the Schwarzschild black hole \cite{c7}, which is given by:
\begin{align}
\Omega_\phi = \sqrt{\frac{M}{r^3}}.
\end{align}
In the limiting case $\alpha \to 0$ or $\beta \to 1/2$, the angular velocity simplifies to an expression identical to that of the Reissner–Nordstr\"om black hole, and is given by:
\begin{align}
\Omega_\phi = \sqrt{\frac{M}{r^3}-\frac{q^2}{r^4}}.
\end{align}
To express the angular frequency in terms of the physical frequency measured in Hertz (Hz), the following conversion relation is utilized:

\begin{align}
    \nu_\phi = \frac{c^3}{2\pi G M} \cdot \sqrt{\frac{M}{r^3}+\frac{3 \alpha  (2 \beta -1) q^4}{10 r^8}-\frac{q^2}{r^4}}
\end{align}
\begin{figure}[!h]
\centering
\includegraphics[width=0.95\linewidth]{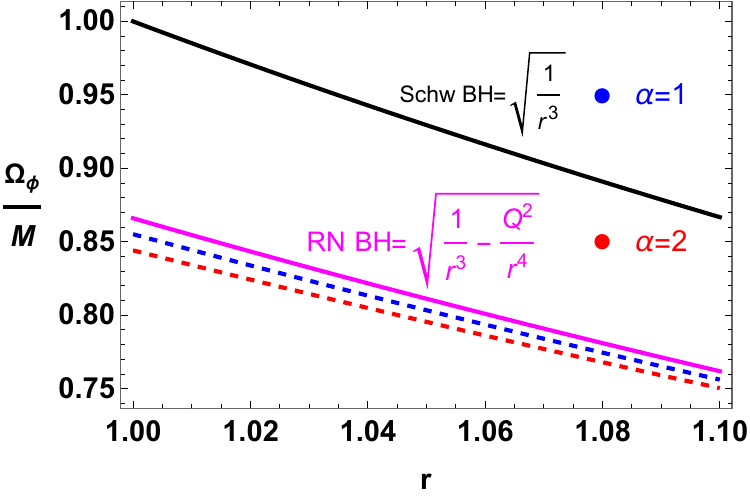}
\vspace{-0.3cm}
\caption{Angular frequency $\Omega_\phi/M$ comparison for Schwarzschild, RN, and NED-$f(R,T)$ black holes, showing deviations with increasing nonlinearity parameter $\alpha$} 
\label{fig7}
\end{figure}
\begin{figure}[!h]
\centering
\includegraphics[width=0.95\linewidth]{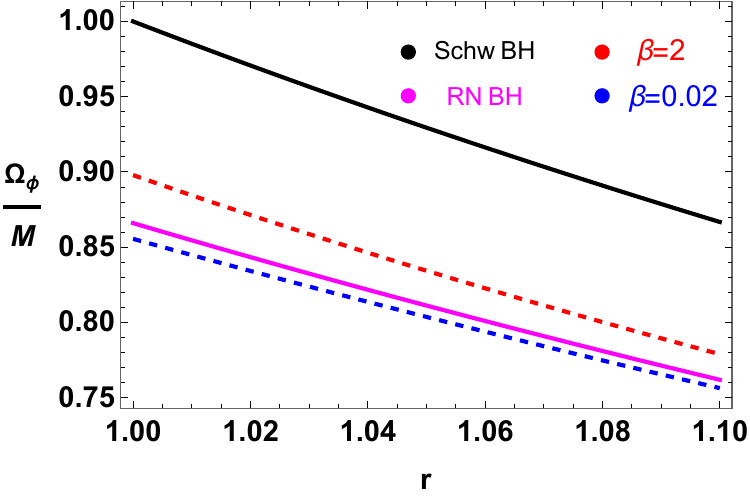}
\vspace{-0.3cm}
\caption{Angular frequency $\Omega_\phi/M$ comparison for Schwarzschild, RN, and NED-$f(R,T)$ black holes, showing deviations with increasing coupling parameter $\beta$} 
\label{fig8}
\end{figure}
Figures \ref{fig7} and \ref{fig8} depict the radial dependence of the Keplerian frequency $\Omega_\phi/M$ as a function of the radial coordinate $r$. The black solid line denotes the Schwarzschild black hole case, whereas the magenta curve corresponds to the Reissner–Nordstr\"om (RN) black hole solution.
  In Figure \ref{fig7}, the blue and red dashed curves represent the behavior of NED black holes within the framework of $f(R,T)$ gravity for different values of the nonlinearity parameter, with $\alpha = 1$ and $\alpha = 2$, respectively. Similarly, in Figure \ref{fig8}, the blue and red dashed lines correspond to black holes with fixed $\alpha$ and varying coupling parameter $\beta = 0.02$ and $\beta = 2$, respectively.\\
 In Figure \ref{fig7} and \ref{fig8}, it is clearly observed that the Keplerian frequency $\Omega_\phi$ diminishes with increasing radial distance $r$. Furthermore, the presence of electric charge and nonlinear electrodynamics (NED) corrections results in a noticeable suppression of $\Omega_\phi$ relative to the Schwarzschild black hole scenario.
 In Figure \ref{fig7}, an increase in the nonlinearity parameter $\alpha$ results in a further suppression of the orbital frequency, underscoring the influential role of the NED parameter in modifying the dynamical trajectories of test particles. In contrast, Figure \ref{fig8} reveals an opposite trend: as the matter-curvature coupling parameter $\beta$ increases, the orbital frequency also increases. This contrasting behavior highlights the distinct impacts of electromagnetic nonlinearity and curvature-matter coupling in the $f(R,T)$ gravity framework, where $\alpha$ tends to weaken the gravitational pull experienced by the particle, while higher $\beta$ values enhance it, leading to elevated orbital velocities.\\
 In Figure \ref{fig7}, it is observed that as the parameter $\alpha$ increases, the angular frequency profile progressively diverges away from that of the Schwarzschild and Reissner–Nordstr\"om (RN) black holes. Conversely, with decreasing $\alpha$, the profile gradually approaches that of the RN black hole, indicating closer agreement with the charged solution than with the neutral Schwarzschild case.
In Figure \ref{fig8}, it is observed that as the coupling parameter $\beta$ increases, the angular frequency profile progressively deviates from that of the Reissner–Nordstr\"om (RN) black holes and towards the Schwarzschild case. Conversely, for smaller values of $\beta$, the angular frequency curve closely aligns with the RN case, indicating that weaker matter-curvature coupling leads to dynamics more consistent with standard charged black hole solutions.

\begin{figure*}[t!]
\centering
\includegraphics[width=0.3\linewidth]{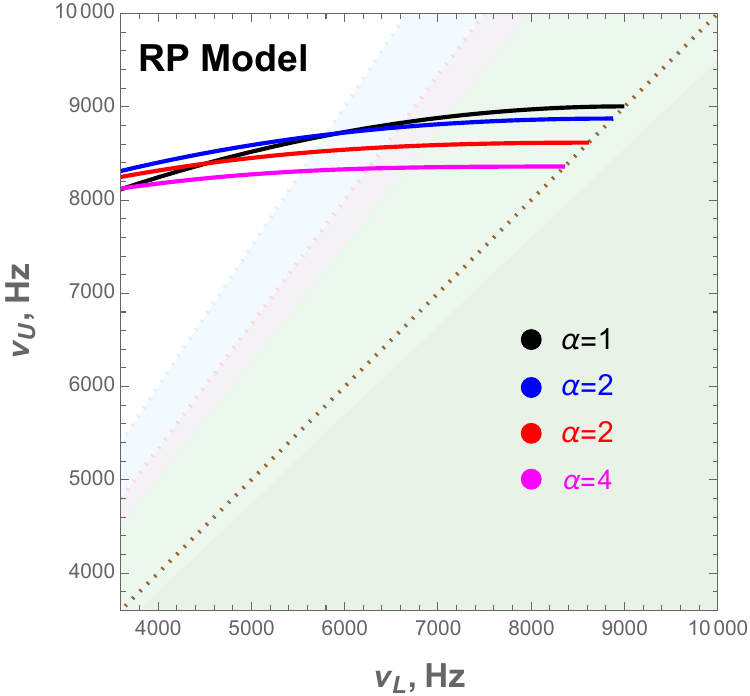}
\includegraphics[width=0.3\linewidth]{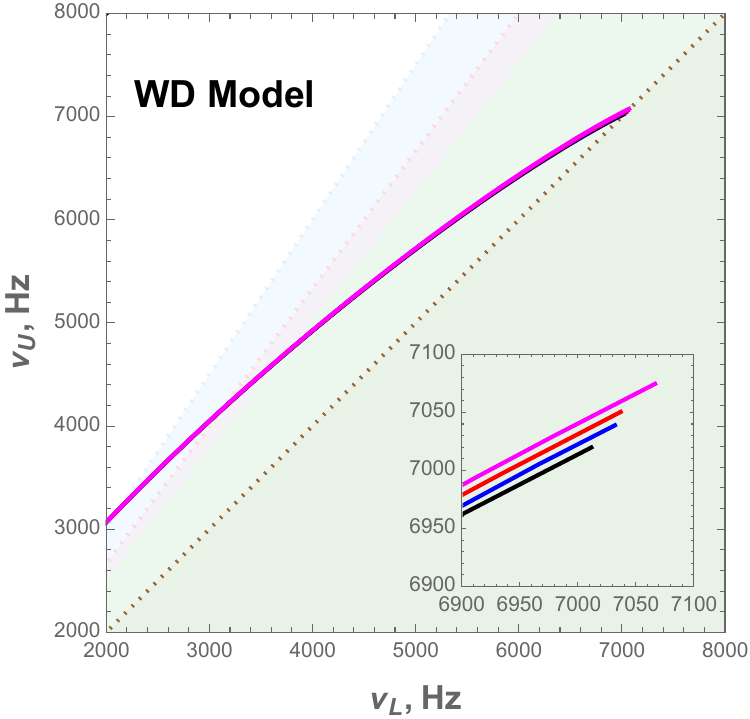}
\includegraphics[width=0.3\linewidth]{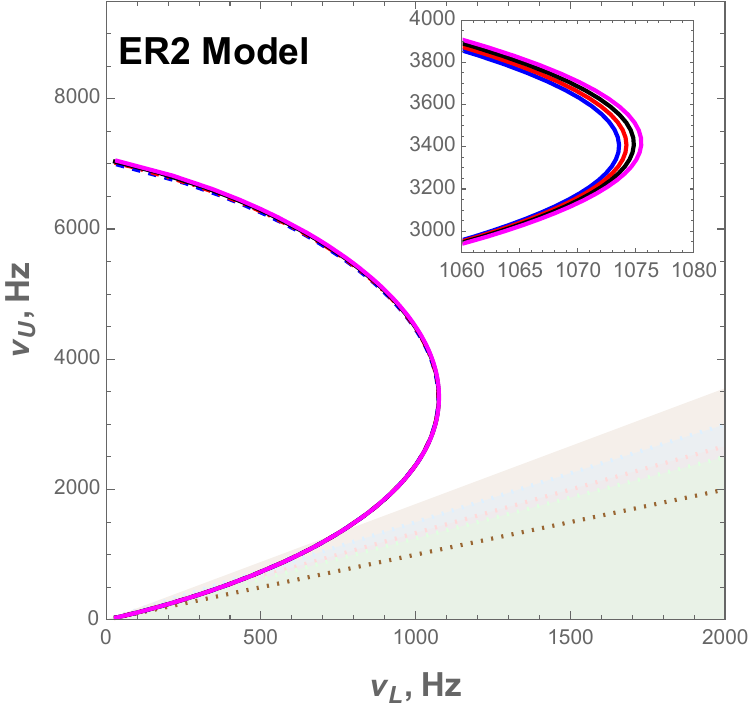}
\includegraphics[width=0.3\linewidth]{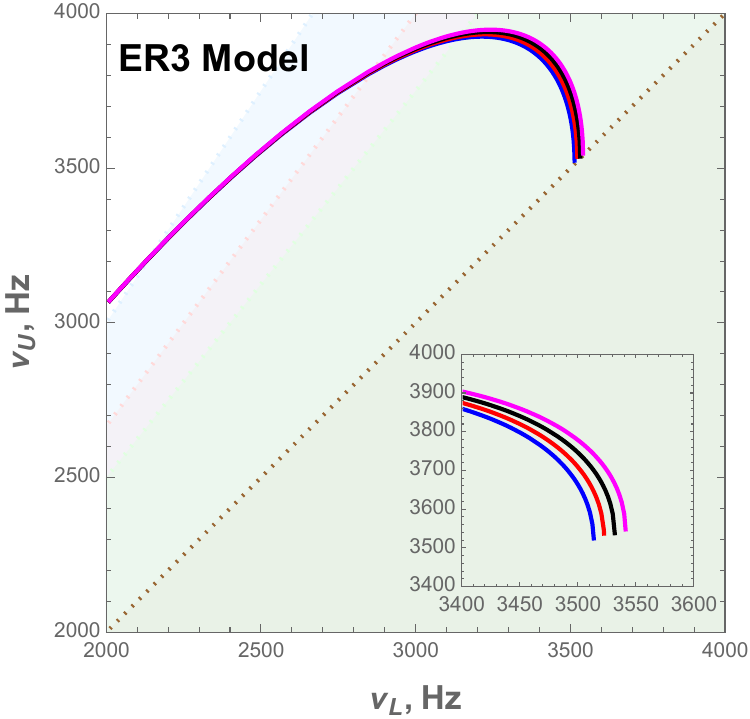}
\includegraphics[width=0.3\linewidth]{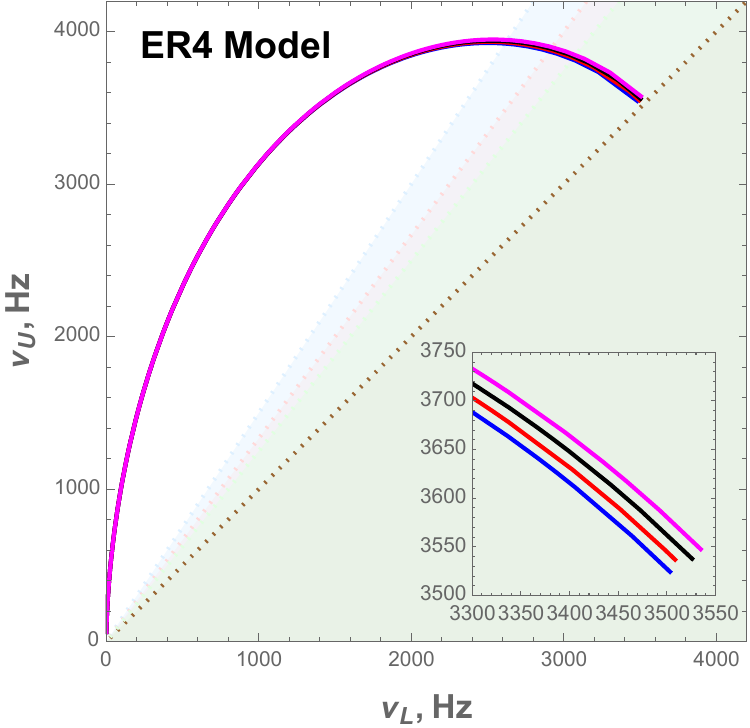}
\caption{Relation between upper ($\nu_U$) and lower ($\nu_L$) twin-peak QPO frequencies for the RP, WD, and ER2–ER4 models in the spacetime of NED black holes within $f(R,T)$ gravity, with curves corresponding to varying values of the nonlinearity parameter $\alpha$.
} 
\label{fig9}
\end{figure*}
\begin{figure*}[t!]
\centering
\includegraphics[width=0.3\linewidth]{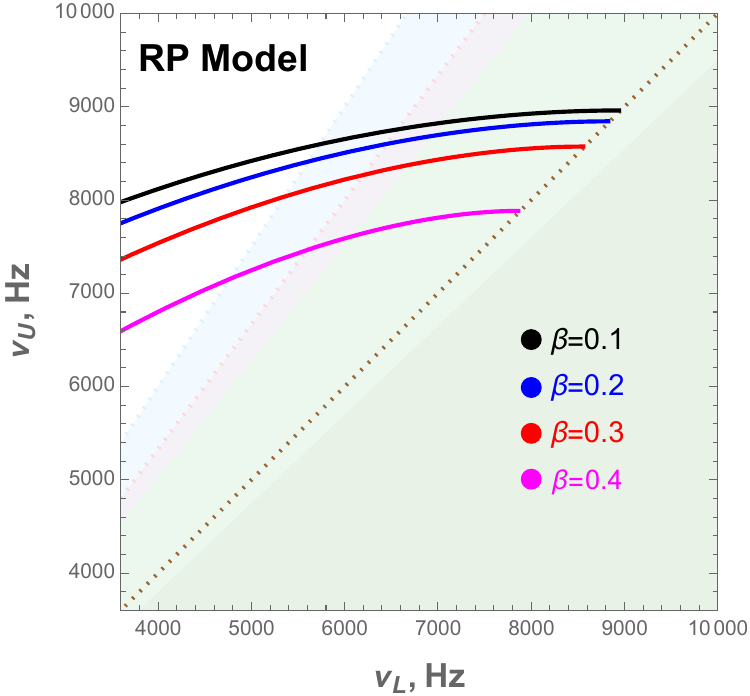}
\includegraphics[width=0.3\linewidth]{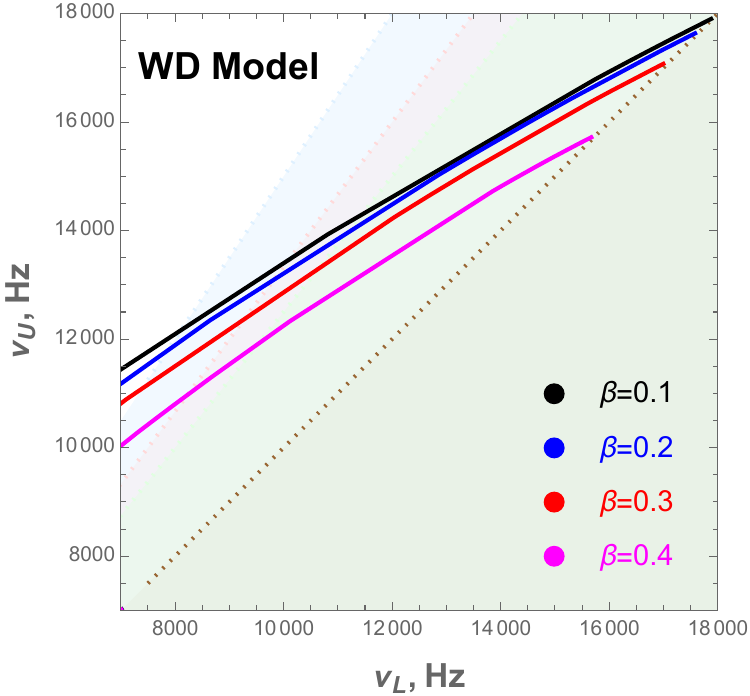}
\includegraphics[width=0.3\linewidth]{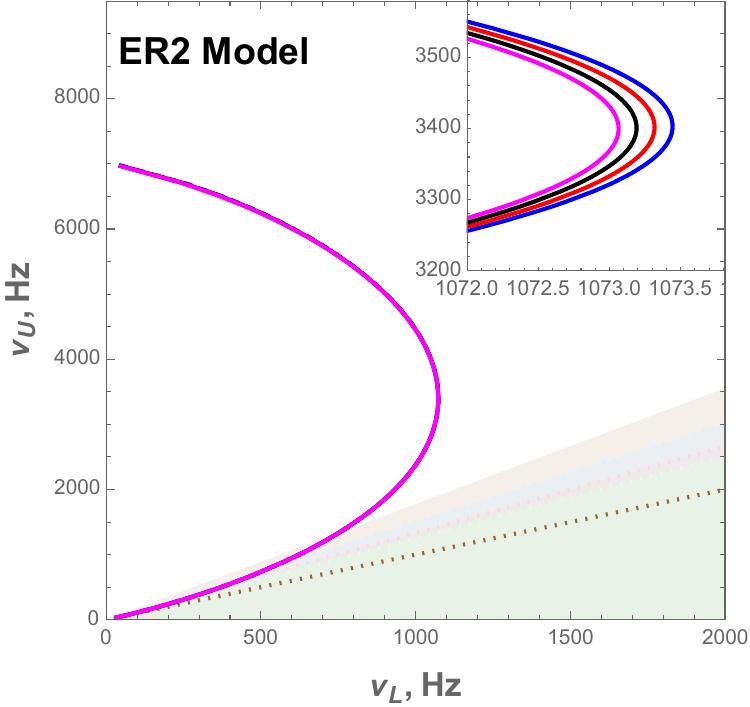}
\includegraphics[width=0.3\linewidth]{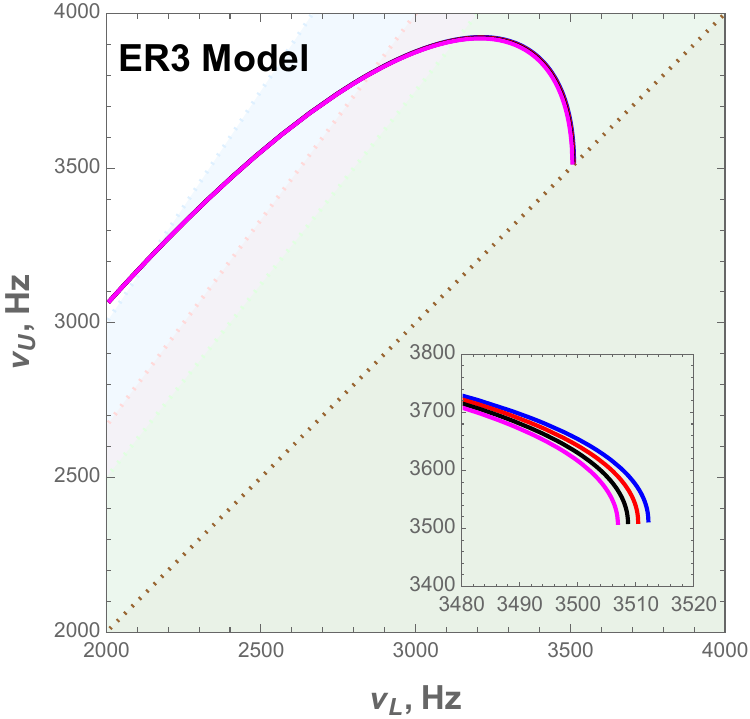}
\includegraphics[width=0.3\linewidth]{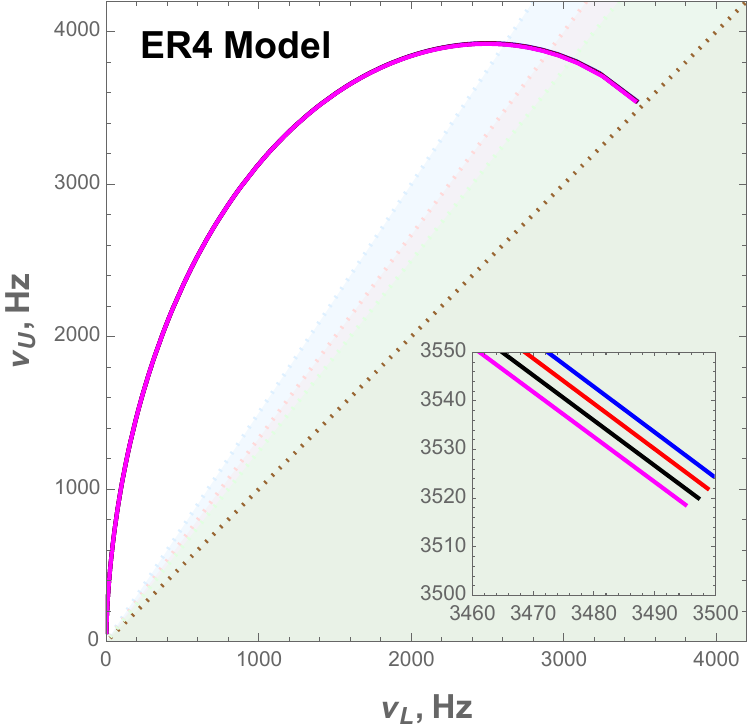}
\caption{Relation between upper ($\nu_U$) and lower ($\nu_L$) twin-peak QPO frequencies for the RP, WD, and ER2–ER4 models in the spacetime of NED black holes within $f(R,T)$ gravity, with curves corresponding to varying values of the coupling parameter $\beta$.
} 
\label{fig10}
\end{figure*}

\subsection{Harmonic oscillations}

In this subsection, we analyze the fundamental frequencies corresponding to the oscillatory motion of test particles around a black hole governed by nonlinear electrodynamics (NED) in the framework of $f(R,T)$ gravity. These characteristic frequencies which are namely the radial and vertical (latitudinal) components are derived by introducing small perturbations around a stable circular orbit, expressed as $r \rightarrow r_0 + \delta r$ and $\theta \rightarrow \theta_0 + \delta \theta$. The effective potential $V_{\text{eff}}(r, \theta)$ may be expanded in a Taylor series about the circular orbit position $(r_0, \theta_0)$, yielding the following expression:

\begin{align}
V_{\text{eff}}(r, \theta) &= V_{\text{eff}}(r_0, \theta_0) + \delta r \left. \frac{\partial V_{\text{eff}}}{\partial r} \right|_{r_0, \theta_0} + \delta \theta \left. \frac{\partial V_{\text{eff}}}{\partial \theta} \right|_{r_0, \theta_0} \nonumber \\
&+ \frac{1}{2} \delta r^2 \left. \frac{\partial^2 V_{\text{eff}}}{\partial r^2} \right|_{r_0, \theta_0} + \frac{1}{2} \delta \theta^2 \left. \frac{\partial^2 V_{\text{eff}}}{\partial \theta^2} \right|_{r_0, \theta_0} \nonumber \\
&+ \delta r \delta \theta \left. \frac{\partial^2 V_{\text{eff}}}{\partial r \partial \theta} \right|_{r_0, \theta_0} + \mathcal{O}(\delta r^3, \delta \theta^3).
\label{exp}
\end{align}
By imposing the criteria for circular motion and stability, only the second-order partial derivatives of the effective potential remain relevant. This results in harmonic oscillator equations governing the radial and vertical perturbations within the equatorial plane, as perceived by a distant observer as \cite{Bambi2017book}:

\begin{align}
    \frac{d^2\delta r}{dt^2}+\Omega^2_r\delta r=0,\  \frac{d^2\delta\theta }{dt^2}+\Omega^2_\theta \delta\theta=0,
\end{align}
where
\begin{align}
 \Omega_r^2=-\frac{1}{2g_{rr}\Dot{t}^2}\partial_r^2 V_\text{eff}(r,\theta)\Big\arrowvert_{\theta=\pi/2},
\end{align}
\begin{align}
        \Omega_\theta^2=-\frac{1}{2g_{\theta\theta}\Dot{t}^2}\partial_\theta^2 V_\text{eff}(r,\theta)\Big\arrowvert_{\theta=\pi/2},
\end{align}
 represent the frequencies associated with radial and vertical (latitudinal) oscillations, respectively. For black holes arising from nonlinear electrodynamics (NED) within the framework of $f(R,T)$ gravity, the corresponding expressions for these characteristic oscillation frequencies are given by:
\begin{align}
 \Omega_r^2=\text{See appendix \ref{A}},
\end{align}
\begin{align}
        \Omega_\theta^2= \Omega_\phi^2=\frac{M}{r^3}+\frac{3 \alpha  (2 \beta -1) q^4}{10 r^8}-\frac{q^2}{r^4}
\end{align}
To express these oscillation frequencies in physical units of Hertz (Hz), the following conversion formula is utilized:
\begin{align}
    \nu_i = \frac{c^3}{2\pi G M} \cdot \Omega_i
\end{align}

\section{QPO Models and QPO orbits}
\subsection{QPO Models}
\begin{figure*}[t!]
\centering
\includegraphics[width=0.3\linewidth]{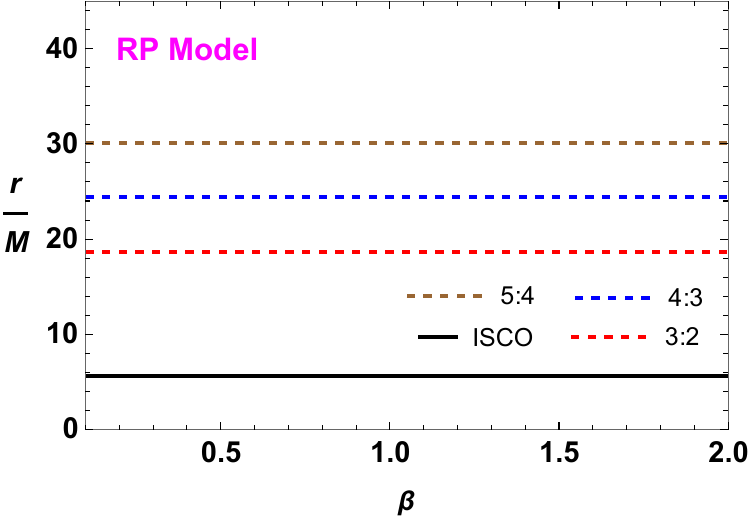}
\includegraphics[width=0.3\linewidth]{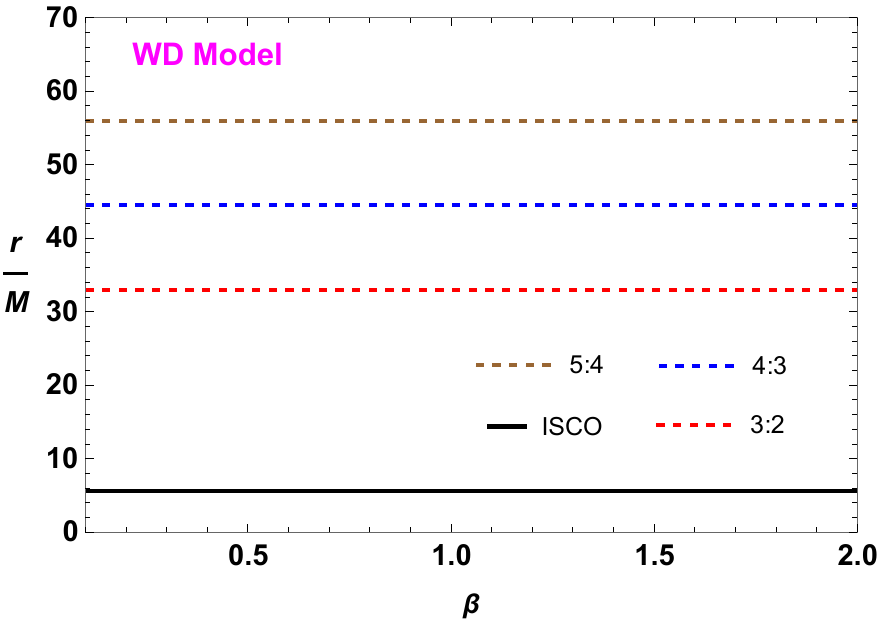}
\includegraphics[width=0.3\linewidth]{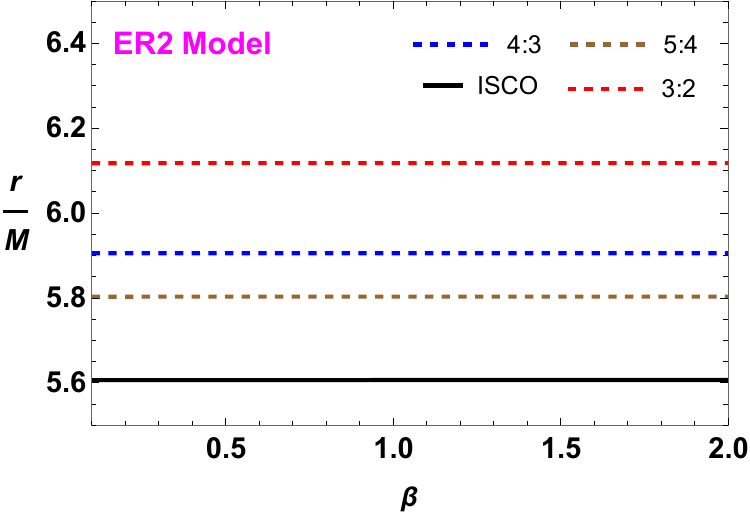}
\includegraphics[width=0.3\linewidth]{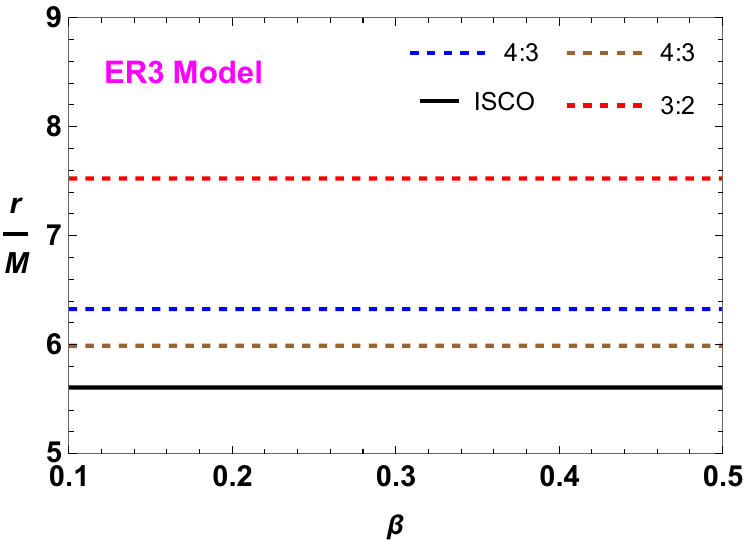}
\includegraphics[width=0.3\linewidth]{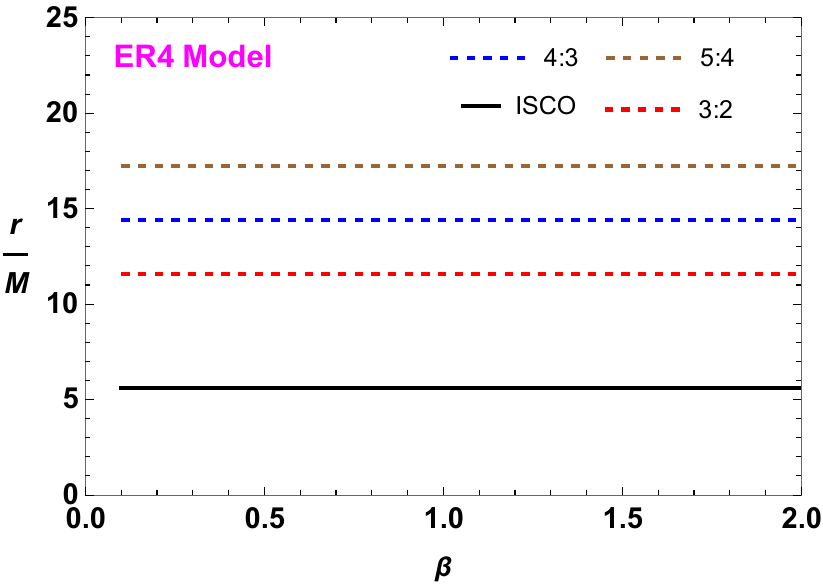}
\caption{ Radial locations of QPO-generating orbits as functions of the coupling parameter $\beta$ across the  RP, WD, and ER2–ER4 quasi-periodic oscillation models.
} 
\label{fig11}
\end{figure*}
In this section, we explore the characteristics of twin-peak quasi-periodic oscillations (QPOs) in the spacetime of a black hole described by nonlinear electrodynamics (NED) within the framework of $f(R,T)$ gravity, and compare the results with those obtained for Schwarzschild and Reissner–Nordstr\"om (RN) black holes. The upper ($\nu_U$) and lower ($\nu_L$) QPO frequencies are expressed as functions of the radial coordinate and black hole parameters, in accordance with several well-established QPO models~\cite{51}. The analysis incorporates the following QPO frameworks~\cite{51}:

\begin{itemize}
    \item \textbf{Relativistic Precession (RP) model:} 
    The relativistic precession (RP) model attributes the quasi-periodic oscillations (QPOs) observed in X-ray binary systems to the intrinsic motion of matter within the curved spacetime surrounding a compact object, interpreting these oscillations as a fundamental manifestation of general relativistic effects in strong gravity regimes.
 Within this framework, localized plasma inhomogeneities within the accretion disk are modeled as following mildly eccentric and inclined geodesic trajectories in the vicinity of the black hole. These slight deviations from perfectly circular orbits generate distinct frequencies that are intrinsically linked to the underlying orbital dynamics.
In the context of the relativistic precession (RP) model, high-frequency quasi-periodic oscillations (QPOs) are attributed to the fundamental coordinate frequencies arising from geodesic motion. Specifically, the upper kHz QPO is associated with the orbital (Keplerian) frequency, $\nu_{U} = \nu_{\phi}$, while the lower kHz QPO corresponds to the periastron precession frequency, defined as $\nu_{L} = \nu_{\phi} - \nu_{r}$, where $\nu_{r}$ denotes the radial epicyclic frequency.
 Moreover, the low-frequency QPOs detected in black hole systems are commonly attributed to the nodal precession frequency, given by $\nu_{\phi} - \nu_{\theta}$, which arises from vertical oscillations induced by the frame-dragging effect in curved spacetime. Importantly, this frequency diminishes to zero in the Schwarzschild limit, where the vertical epicyclic frequency $\nu_{\theta}$ becomes equal to the orbital frequency $\nu_{\phi}$, indicating the absence of nodal precession in non-rotating spacetimes.
These frequency associations are well-supported by general relativistic modifications to orbital motion, such as frame dragging and spacetime curvature, which inherently give rise to the observed precessional effects.
 The detection of harmonic structures in both neutron star and black hole systems further reinforces this interpretation, particularly in instances where even multiples of the nodal precession frequency prevail in the power spectrum \cite{StellaVietri1998, StellaVietri1999, MorsinkStella1999}. The RP model thereby provides a straightforward, geometrically grounded explanation for QPOs, based purely on relativistic effects without requiring resonance phenomena or strong magnetic fields.

    \item \textbf{Epicyclic Resonance (ER) models:} Under the assumption of a geometrically thick accretion disk, the resonance conditions yield the following expressions for the QPO frequencies:
 \begin{itemize}
        \item ER2: $\nu_U = 2\nu_\theta - \nu_r$, $\nu_L = \nu_r$,
        \item ER3: $\nu_U = \nu_\theta + \nu_r$, $\nu_L = \nu_\theta$,
        \item ER4: $\nu_U = \nu_\theta + \nu_r$, $\nu_L = \nu_\theta - \nu_r$.
    \end{itemize}
These configurations represent different resonant interactions between the vertical and radial epicyclic oscillations in the disk.

    \item \textbf{Warped Disk (WD) model:} This model, which is based on a non-conventional geometrical configuration of the accretion disk \cite{wd1,wd2}, interprets high-frequency QPOs (HFQPOs) as a consequence of nonlinear resonant interactions between relativistically distorted, warped disk structures and intrinsic oscillation modes of the disk.
 These resonant interactions encompass both radial and vertical components: radial (horizontal) resonances can excite both g-mode and p-mode oscillations, whereas vertical resonances predominantly stimulate g-mode oscillations exclusively \cite{wd2}.
The underlying mechanism driving these resonances is associated with the non-monotonic radial profile of the epicyclic frequency as a function of the coordinate $r$. Within this theoretical framework, the upper high-frequency QPO is interpreted as \cite{wd1,wd2}: 
\begin{equation}\nu_U = 2\nu_\phi - \nu_r, \nu_L = 2(\nu_\phi - \nu_r)\end{equation}

\end{itemize}
Figure \ref{fig9} and \ref{fig10} presents the computed correlations between $\nu_U$ and $\nu_L$ for black holes described by nonlinear electrodynamics (NED) within the framework of $f(R,T)$ gravity, evaluated across different QPO models and for various values of the nonlinearity parameter $\alpha$ and coupling parameter $\beta$ respectively. The plot includes reference lines representing characteristic frequency ratios of $3\!:\!2$, $4\!:\!3$, $5\!:\!4$, and $1\!:\!1$, the last of which signifies scenarios where the upper and lower QPOs coalesce into a single peak commonly referred to as the "QPO graveyard."

\subsection{QPO orbits}
In this subsection, we investigate the influence of the parameter $\beta$ on the orbital radii at which quasi-periodic oscillations (QPOs) with characteristic frequency ratios such as $3\!:\!2$, $4\!:\!3$, and $5\!:\!4$ can emerge across the various QPO models considered. These particular radii are obtained by solving the corresponding resonance conditions.
\begin{equation}
a\,\nu_U(M, r, \ell) = b\,\nu_L(M, r, \ell), \label{resonance}
\end{equation}
Here, $a$ and $b$ are integers denoting the specific resonance ratio. For the RP, WD, and ER2–ER4 models, the associated resonance condition can be solved numerically to obtain the radial coordinate $r$ for various values of the coupling parameter $\beta$. The resulting numerical solutions are presented in Figure~\ref{fig11}.

Figure~\ref{fig11} displays the dependence of the QPO-generating orbital radius, normalized by $r/M$, on the coupling parameter $\beta$, across all five QPO models considered: RP, WD, ER2, ER3, and ER4.
Each panel illustrates the radial positions at which the resonant frequency ratios $3\!:\!2$, $4\!:\!3$, and $5\!:\!4$ occur, with the innermost stable circular orbit (ISCO) radius indicated by a solid black curve.
 It is apparent that the WD model yields QPOs at substantially larger orbital radii relative to the other models, signifying the largest QPO-generating distances among the five frameworks considered.
 The RP model yields the next largest QPO-generating radii, whereas the ER2 model exhibits the most compact resonance orbits among all models considered. Furthermore, for each model, the orbital radius associated with a given resonance varies non-monotonically with the coupling parameter $\beta$, reflecting the intricate influence of matter-curvature coupling on the location of resonance orbits in the modified $f(R,T)$ gravity framework.

{\LARGE
\begin{table*}[htb]
\centering
\begin{tabular}{|c|c|c|c|}
\hline
\textbf{Source} & \textbf{Mass} (in $M_\odot$) & \textbf{Upper Frequency (Hz)} & \textbf{Lower Frequency (Hz)} \\
\hline
GRO J1655$-$40 & $5.4 \pm 0.3$ \cite{t57} & $441 \pm 2$ \cite{t58} & $298 \pm 4$ \cite{t58} \\
\hline
XTE J1550$-$564 & $9.1 \pm 0.61$ \cite{t59} & $276 \pm 3$ & $184 \pm 5$ \\
\hline
GRS 1915$+$105 & $12.4^{+2.0}_{-1.8}$ \cite{t60} & $168 \pm 3$ & $113 \pm 5$ \\
\hline
H 1743$+$322 & $8.0 - 14.07$ \cite{t61,t62,t63} & $242 \pm 3$ & $166 \pm 5$ \\
\hline
M82 X-1 & $415 \pm 63$ \cite{nature} & $5.07 \pm 0.06$\cite{nature}  & $3.32 \pm 0.06$\cite{nature}  \\
\hline
Sgr A$^*$ & $(3.5 - 4.9) \times 10^6$ \cite{t64,t65} & $(1.445 \pm 0.16) \times 10^{-3}$\cite{t66} & $(0.886 \pm 0.04) \times 10^{-3}$\cite{t66} \\
\hline
\end{tabular}
\caption{Observational QPO data for different Black hole sources with estimated mass  \cite{52}}
\label{tab1}
\end{table*}
}
%
%\begin{table*}[htb]
%\centering
%\setlength{\tabcolsep}{12pt} % wider column spacing
%\renewcommand{\arraystretch}{1.4} % taller rows
%\begin{tabular}{|c|c|c|c|c|}
%\hline
%\textbf{Source} & \boldmath{$M$} & \boldmath{$b$} & \boldmath{$Q/M$} & \boldmath{$r/M$} \\
%\hline
%  \textbf{GRO J1655-40}     & $5.75 \pm 0.37$                  & $1.06^{+0.63}_{-0.94}$ & $0.24^{+0.13}_{-0.20}$ & $5.45^{+0.23}_{-0.27}$ \\
%\hline
%\textbf{XTE J1550-564}   & $9.9 \pm 1.1$                    & $1.15^{+0.69}_{-0.62}$ & $0.39 \pm 0.19$        & $5.17^{+0.39}_{-0.45}$ \\

%\hline
% \textbf{GRS 1915+105}   & $14.37^{+0.57}_{-0.33}$          & $1.38 \pm 0.87$        & $0.28^{+0.15}_{-0.17}$ & $5.60^{+0.12}_{-0.16}$ \\
%\hline
%\textbf{H 1743+322}     & $12.6 \pm 1.3$                   & $1.35^{+0.76}_{-0.88}$ & $0.42^{+0.24}_{-0.31}$ & $4.76^{+0.38}_{-0.43}$ \\
%\hline
%\textbf{M82 X-1}  & $407^{+0.80}_{-1.00}$  & $1.008 \pm 0.30$    & $0.084^{+0.041}_{-0.072}$ & $3.0091^{+0.0052}_{-0.0080}$ \\
%\hline
%\textbf{Sgr A*}  & $(4.17^{+0.39}_{-0.46}) \times 10^6$ & $1.53 \pm 0.85$    & $0.54^{+0.32}_{-0.27}$ & $4.9^{+1.2}_{-1.7}$ \\
%\hline

%\end{tabular}
%\caption{Posterior estimates of the parameters \(M\), \(b\), \(Q/M\), and \(r/M\) obtained from MCMC analysis.}
%\label{tab2}
%\end{table*}

\section{Monte Carlo Markov chain (MCMC) analysis}

\begin{figure*}[t]
    \centering
    \subfloat[GRO J1655-40]{%
        \includegraphics[width=0.45\textwidth]{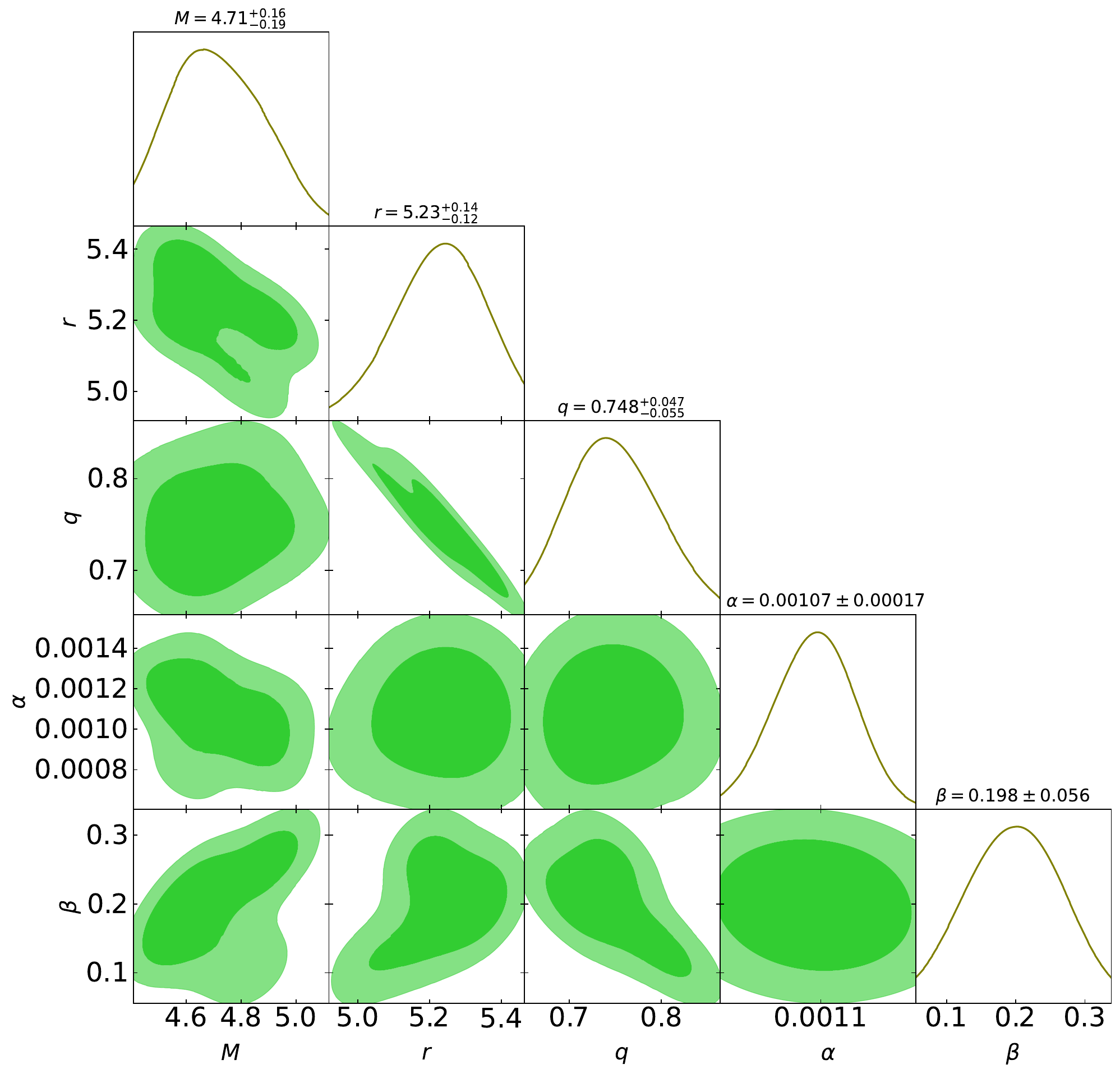}
    }\hfill
    \subfloat[XTE J1550-564]{%
        \includegraphics[width=0.45\textwidth]{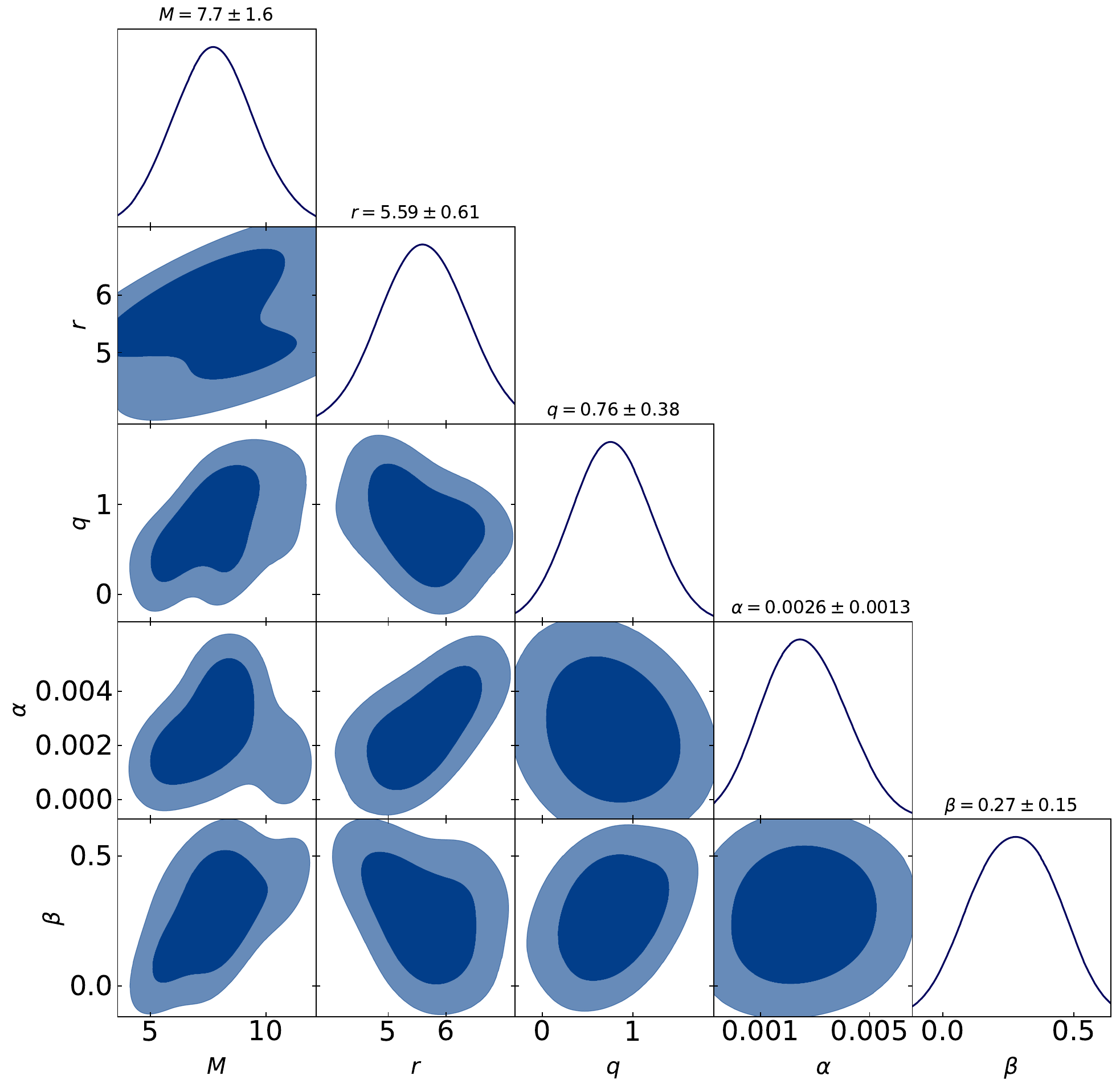}
    }\hfill
    \subfloat[GRS 1915+105]{%
        \includegraphics[width=0.45\textwidth]{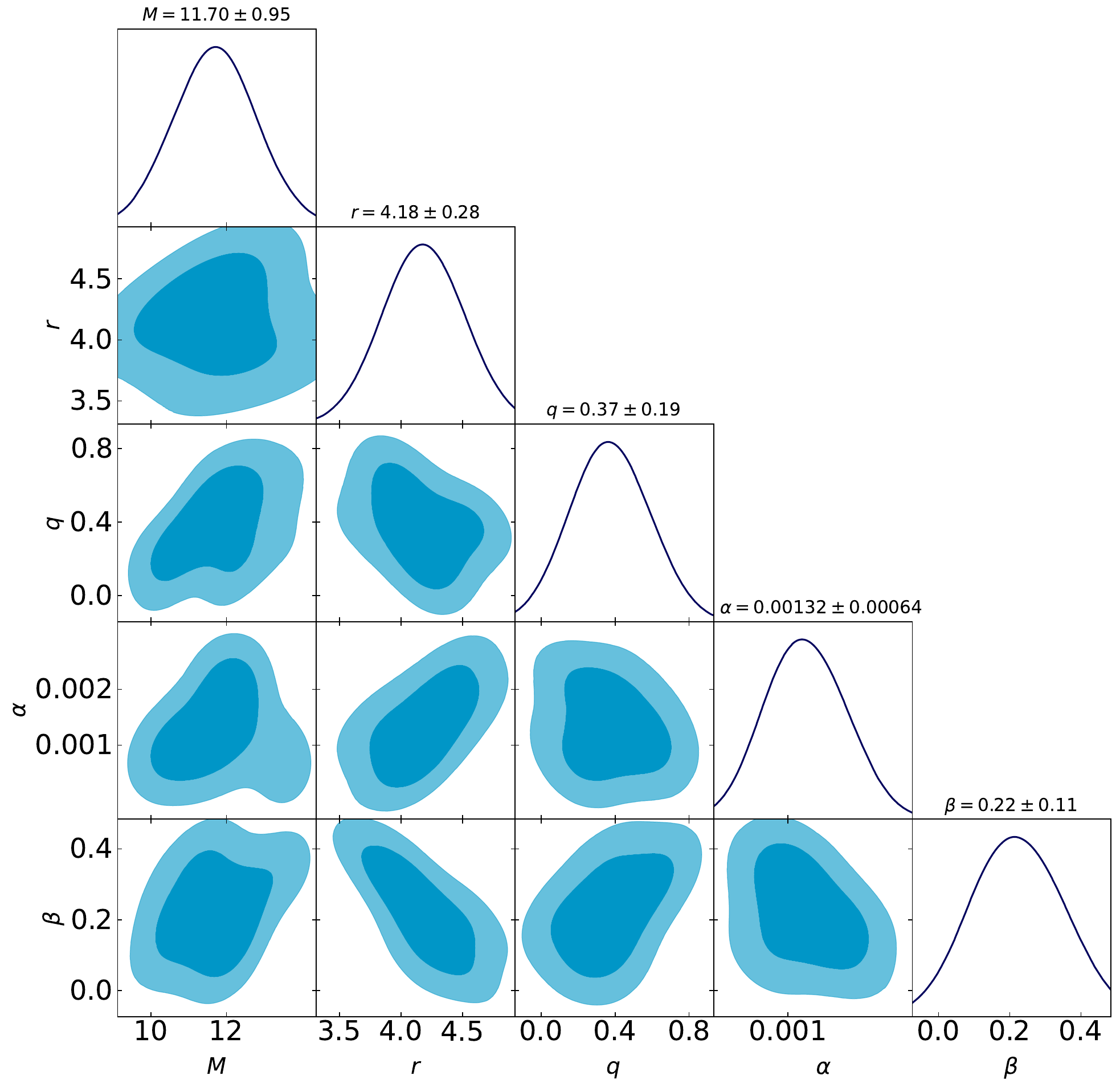}
    }\hfill
    \subfloat[H 1743+322]{%
        \includegraphics[width=0.45\textwidth]{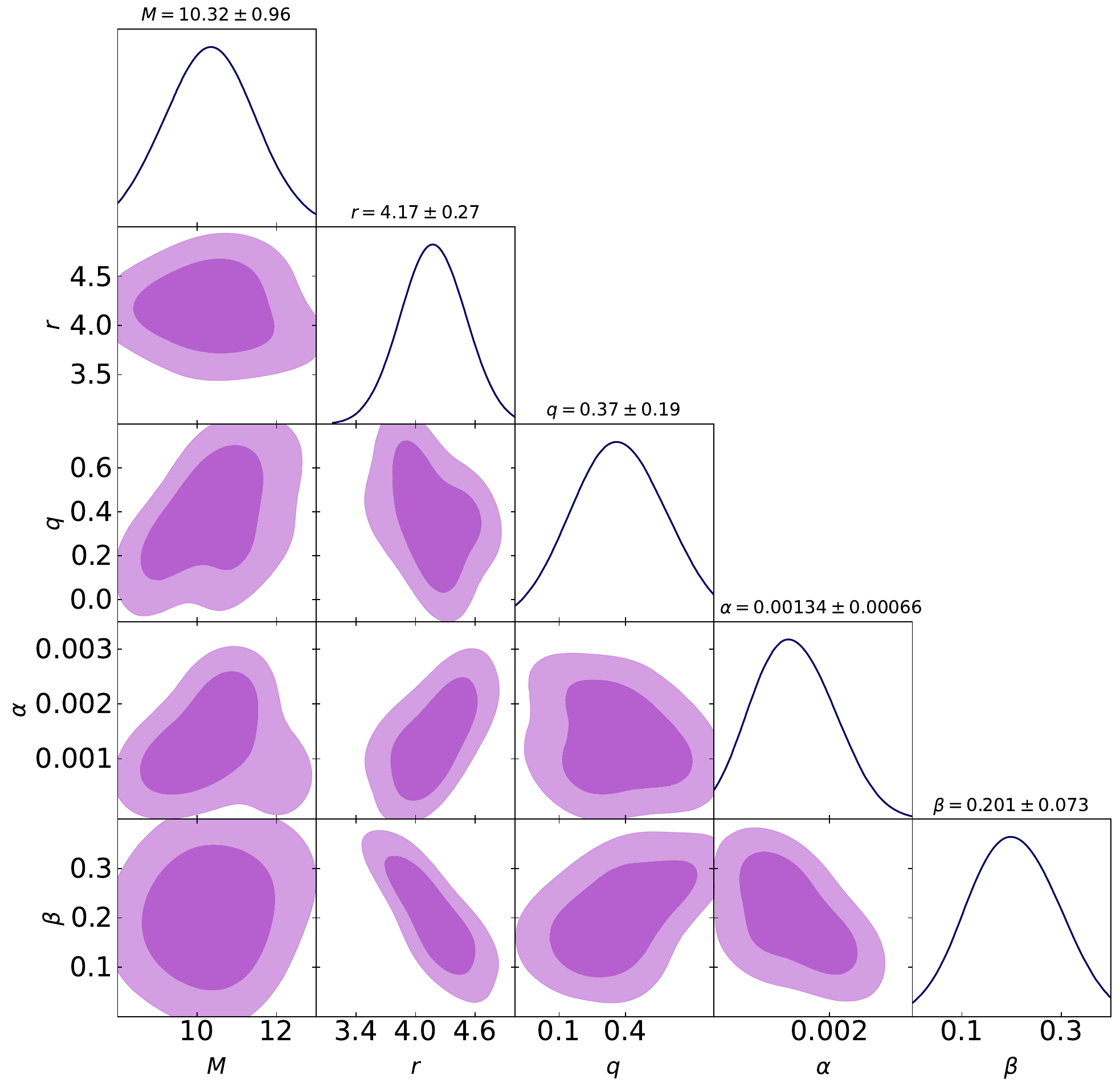}
    }\hfill
    \caption{Corner plots showing posterior distributions for the parameters \(M\), \(b\), \(Q/M\), and \(r/M\) obtained from the MCMC analysis for each black hole source.}
    \label{fig12}
\end{figure*}

In this section, we conduct a Markov Chain Monte Carlo (MCMC) analysis to constrain the model parameters where specifically the nonlinearity parameter $\alpha$ associated with the NED sector and the coupling parameter $\beta$ from the $f(R,T)$ gravity framework using observational data from six well established black hole sources spanning three distinct mass regimes: GRO J1655–40, XTE J1550–564, GRS 1915+105, H 1743–322, M82 X-1, and Sgr A*.
 Among the selected sources, GRO J1655–40, XTE J1550–564, GRS 1915+105, and H 1743–322 are categorized as stellar-mass black holes, M82 X-1 serves as a representative of the intermediate-mass black hole class, and Sgr A* corresponds to a supermassive black hole. A summary of these black hole systems along with their relevant observational properties is provided in Table~\ref{tab1}.
  \\
In this study, we employed the relativistic precession (RP) model as a representative framework to exemplify the implementation of MCMC methods for constraining black hole parameters. This selection was made not out of a specific preference for the RP model, but solely to demonstrate the analytical approach and methodology.
However, motivated by observational evidence indicating that QPO frequencies frequently adhere to the ratio $\nu_U / \nu_L \approx 3\!:\!2$, we have adopted the 3:2 frequency configuration as the basis for our analysis.
\\
 The Bayesian posterior distribution is expressed as follows:

\begin{equation}
P(\boldsymbol{\theta} | D, M) = \frac{P(D | \boldsymbol{\theta}, M)\, \pi(\boldsymbol{\theta} | M)}{P(D | M)},
\label{eq:posterior}
\end{equation}
Here, $\pi(\boldsymbol{\theta})$ represents the prior probability distribution over the parameter set $\boldsymbol{\theta} = \{M, \alpha, \frac{q}{M}, \frac{r}{M}\}$, while $P(D | \boldsymbol{\theta}, M)$ denotes the likelihood function based on the observational data $D$. Gaussian priors are assumed for each parameter, specified as follows:

\begin{equation}
\pi(\theta_i) \propto \exp\left(-\frac{1}{2} \left(\frac{\theta_i - \theta_{0,i}}{\sigma_i} \right)^2\right), \quad \theta_{\mathrm{low},i} < \theta_i < \theta_{\mathrm{high},i},
\end{equation}
Here, $\theta_{0,i}$ and $\sigma_i$ denote the mean and standard deviation of the parameters as reported in the literature, while the imposed bounds guarantee the physical admissibility of the parameter space.
\\
The likelihood function accounts for contributions from both the upper and lower quasi-periodic oscillation (QPO) frequencies, ensuring consistency with observational constraints as:

\begin{equation}
\log \mathcal{L} = \log \mathcal{L}_U + \log \mathcal{L}_L,
\end{equation}
with
\begin{equation}
\log \mathcal{L}_U = -\frac{1}{2} \sum_i \frac{(\nu^{\mathrm{obs}}_{\phi,i} - \nu^{\mathrm{th}}_{\phi,i})^2}{(\sigma^{\mathrm{obs}}_{\phi,i})^2},
\end{equation}
\begin{equation}
\log \mathcal{L}_L = -\frac{1}{2} \sum_i \frac{(\nu^{\mathrm{obs}}_{L,i} - \nu^{\mathrm{th}}_{L,i})^2}{(\sigma^{\mathrm{obs}}_{L,i})^2},
\end{equation}
In this analysis, $\nu^{\mathrm{obs}}_{\phi,i}$ and $\nu^{\mathrm{obs}}_{L,i}$ represent the observed orbital and lower quasi-periodic oscillation (QPO) frequencies for the $i^\text{th}$ black hole source, while $\nu^{\mathrm{th}}_{\phi,i}$ and $\nu^{\mathrm{th}}_{L,i}$ denote the corresponding theoretical predictions derived using the RP model.
\\
We have applied our methodology to six distinct black hole systems spanning different mass regimes, utilizing observational QPO data as summarized in Table~\ref{tab:all-models}. A Markov Chain Monte Carlo (MCMC) analysis is performed with Gaussian priors to sample $10^5$ realizations per parameter, facilitating an exhaustive exploration of the underlying parameter space. Our goal is to determine the most probable values of $\{M, \alpha, Q/M, r/M\}$ that best fit the observed data across all five QPO models considered—including RP, WD, and ER2–ER4.
\\
Figure~\ref{fig12} displays the corner plots obtained from our MCMC analysis, where the shaded contours denote the 1$\sigma$ (68\%) and 2$\sigma$ (95\%) credible intervals for the posterior probability distributions. The inferred black hole masses encompass a broad range of astrophysical scales from stellar-mass to supermassive black holes .
\\
For the stellar-mass black hole systems, the inferred mass estimates are in good agreement with existing observational constraints: $M = 4.71^{0.16}_{-0.19}\,M_\odot$ for GRO J1655–40, $7.7\pm 1.6\,M_\odot$ for XTE J1550–564, $11.70 \pm 0.95\,M_\odot$ for GRS 1915+105, and $10.7 \pm 1.1\,M_\odot$ for H 1743+322.
\\
 The corresponding estimates for the NED parameter $\alpha$ lie approximately within the range 0.00107 to 0.0026, with moderate associated uncertainties. The charge $q$ spans values between 0.30 and 0.76, while the normalized radial coordinate $r$ is found to range from 4.06 to 5.59. Additionally, the coupling parameter $\beta$ associated with $f(R,T)$ gravity varies between 0.198 and 0.27 with associated uncertainties across the analyzed sources.
\\
The intermediate-mass black hole M82 X-1 yields a well-constrained mass estimate of $M = 410.4^{+9.8}_{-11}\,M_\odot$, along with a small NED parameter $\alpha = 0.00043^{+0.00022}_{-0.00027}$ and a moderate coupling parameter $\beta = 0.228^{+0.10}_{-0.084}$. The black hole carries a relatively high charge, with $q = 0.33 \pm 0.16$, and the normalized radial coordinate is found to be $r = 3.92 \pm 0.21$. These results suggest a weakly charged black hole with minimal contributions from nonlinear electrodynamics and moderate curvature-matter coupling within the $f(R,T)$ gravity framework.
\\
For the supermassive black hole Sgr A*, our analysis yields a mass estimate of $M = (3.910^{+0.030}_{-0.024}) \times 10^6\,M_\odot$, accompanied by a nonlinear electrodynamics (NED) parameter $\alpha = 0.0059^{+0.0022}_{-0.0026}$ and a matter-curvature coupling parameter $\beta = 0.81 \pm 0.10$. The inferred electric charge is relatively large, $q = 0.95 \pm 0.25$, while the normalized radial parameter $r = 5.45 \pm 0.48$ exhibits good agreement with the shadow radius constraints obtained from Keck and VLTI interferometric observations, which restrict the value of $r_{\text{sh}}/M$ within the range $4.55 \lesssim r_{\text{sh}}/M \lesssim 5.22$ at 1$\sigma$ and $4.21 \lesssim r_{\text{sh}}/M \lesssim 5.56$ at 2$\sigma$ confidence level.
\\
The MCMC results indicate that the NED parameter $\alpha$, while consistently nonzero, tends to assume relatively small values typically much less than unity across all black hole mass regimes, including stellar, intermediate, and supermassive classes. Similarly, the coupling parameter $\beta$ associated with the $f(R,T)$ gravity sector remains within a narrow range around $\beta \sim 0.1-0.8$. Despite the small magnitudes, the persistence of nonzero values for both $\alpha$ and $\beta$ across different sources points to a universal, albeit subtle, imprint of nonlinear electrodynamics and matter-curvature coupling in shaping the QPO spectra. These tightly constrained posteriors thus offer significant support for the existence of small but measurable corrections beyond Maxwellian electrodynamics and Einstein gravity in the strong-field regime.

\begin{table*}[htbp]
\centering
\scriptsize % reduces font size
\renewcommand{\arraystretch}{1.4}
\setlength{\tabcolsep}{6pt} % reduces column separation
\caption{Best-fit parameters for different black hole models (RP, WD, ER2, ER3, ER4) across various sources.}
\begin{tabular}{|l|p{2.4cm}|p{2.4cm}|p{2.7cm}|p{2.3cm}|p{2.3cm}|}
\hline
\textbf{Model} & \(\boldsymbol{\alpha}\) & \(\boldsymbol{\beta}\) & \(\boldsymbol{M~(M_\odot)}\) & \(\boldsymbol{Q/M}\) & \(\boldsymbol{r/M}\) \\
\hline
\multicolumn{6}{|c|}{\textbf{RP Model}} \\
\hline
Sgr A* & $0.005645^{+0.005359}_{-0.004095}$ & $0.8043^{+0.1801}_{-0.1957}$ & $3.918^{+2.635\times10^4}_{-5.051\times10^4}\times10^6$ & $0.9466^{+0.5018}_{-0.4724}$ & $5.508^{+0.8195}_{-0.9496}$ \\
GRO & $0.001057^{+0.0003001}_{-0.0002992}$ & $0.1943^{+0.08765}_{-0.09096}$ & $4.669^{+0.2916}_{-0.1531}$ & $0.7434^{+0.08529}_{-0.05308}$ & $5.224^{+0.1536}_{-0.2183}$ \\
XTE & $0.002404^{+0.002731}_{-0.001963}$ & $0.2683^{+0.2342}_{-0.2445}$ & $7.914^{+3.069}_{-2.804}$ & $0.7632^{+0.6884}_{-0.7239}$ & $5.594^{+1.018}_{-1.087}$ \\
GRS & $0.001213^{+0.001314}_{-0.00089}$ & $0.2115^{+0.1829}_{-0.1684}$ & $11.81^{+1.804}_{-1.784}$ & $0.3685^{+0.3538}_{-0.3403}$ & $4.185^{+0.4741}_{-0.5203}$ \\
H1743 & $0.001753^{+0.003889}_{-0.001363}$ & $0.2003^{+0.1144}_{-0.114}$ & $10.82^{+1.645}_{-2.14}$ & $0.2917^{+0.3869}_{-0.2725}$ & $4.086^{+0.5512}_{-0.6706}$ \\
M82 & $0.0003913^{+0.0004753}_{-0.0003135}$ & $0.2328^{+0.09607}_{-0.1484}$ & $409.3^{+21.14}_{-14.61}$ & $0.3173^{+0.2801}_{-0.3007}$ & $3.907^{+0.4314}_{-0.3856}$ \\
\hline
\multicolumn{6}{|c|}{\textbf{WD Model}} \\
\hline
Sgr A* & $0.005959^{+0.006160}_{-0.004906}$ & $0.7433^{+0.2335}_{-0.2707}$ & $3.904^{+4.771\times10^4}_{-4.219\times10^4}\times10^6$ & $0.7607^{+0.7028}_{-0.6430}$ & $6.115^{+0.6528}_{-0.3609}$ \\
GRO & $0.004857^{+0.0002776}_{-0.0004437}$ & $0.3608^{+0.03525}_{-0.01520}$ & $4.517^{+0.03896}_{-0.01617}$ & $0.7216^{+0.1314}_{-0.1993}$ & $5.131^{+0.4380}_{-0.4258}$ \\
XTE & $0.003307^{+0.007890}_{-0.002828}$ & $0.3924^{+0.3512}_{-0.3643}$ & $8.413^{+2.683}_{-3.227}$ & $0.6421^{+0.7847}_{-0.6061}$ & $5.378^{+1.226}_{-1.182}$ \\
GRS & $0.001655^{+0.003893}_{-0.001303}$ & $0.2458^{+0.1502}_{-0.1970}$ & $12.18^{+1.657}_{-2.081}$ & $0.3028^{+0.4071}_{-0.2857}$ & $4.063^{+0.5916}_{-0.6587}$ \\
H1743 & $0.001245^{+0.001358}_{-0.0009221}$ & $0.1824^{+0.1343}_{-0.1060}$ & $10.41^{+1.796}_{-1.792}$ & $0.3672^{+0.3552}_{-0.3389}$ & $4.179^{+0.4609}_{-0.5144}$ \\
M87 & $0.0006617^{+0.001640}_{-0.0005650}$ & $0.2266^{+0.1001}_{-0.1224}$ & $417.5^{+15.13}_{-21.98}$ & $0.2405^{+0.3400}_{-0.2272}$ & $3.874^{+0.4623}_{-0.5218}$ \\
\hline
\multicolumn{6}{|c|}{\textbf{ER2 Model}} \\
\hline
Sgr A* & $0.00125^{+0.002237}_{-0.001148}$ & $0.7614^{+0.2244}_{-0.4513}$ & $4.032^{+0.04469}_{-0.0534}\times10^6$ & $1.094^{+0.3665}_{-0.4112}$ & $6.063^{+0.7684}_{-1.443}$ \\
GRO & $0.0006184^{+0.0008054}_{-0.0005010}$ & $0.05761^{+0.04647}_{-0.01981}$ & $4.510^{+0.04461}_{-0.008764}$ & $0.6560^{+0.1458}_{-0.1928}$ & $5.298^{+0.3666}_{-0.4113}$ \\
XTE & $0.004028^{+0.01185}_{-0.003441}$ & $0.4468^{+0.6219}_{-0.4133}$ & $9.005^{+3.409}_{-3.797}$ & $0.6043^{+0.7989}_{-0.5647}$ & $5.302^{+1.270}_{-1.104}$ \\
GRS & $0.002036^{+0.005854}_{-0.001674}$ & $0.2973^{+0.2297}_{-0.2458}$ & $12.49^{+2.282}_{-2.347}$ & $0.2933^{+0.4048}_{-0.2719}$ & $3.997^{+0.6425}_{-0.5909}$ \\
H1743 & $0.002057^{+0.005926}_{-0.001695}$ & $0.2369^{+0.1284}_{-0.1411}$ & $11.16^{+2.151}_{-2.410}$ & $0.2931^{+0.4050}_{-0.2721}$ & $3.999^{+0.6206}_{-0.5830}$ \\
M87 & $0.0008203^{+0.002418}_{-0.0007071}$ & $0.2558^{+0.1375}_{-0.1451}$ & $420.3^{+20.93}_{-24.54}$ & $0.2315^{+0.3396}_{-0.2166}$ & $3.823^{+0.5005}_{-0.4641}$ \\
\hline
\multicolumn{6}{|c|}{\textbf{ER3 Model}} \\
\hline
Sgr A* & $0.006906^{+0.003184}_{-0.006519}$ & $0.5115^{+0.4754}_{-0.2181}$ & $4.001^{+5.242\times10^4}_{-7.126\times10^4}\times10^6$ & $0.5239^{+0.9279}_{-0.3985}$ & $6.282^{+0.6575}_{-0.3852}$ \\
GRO & $0.002148^{+0.001392}_{-0.001245}$ & $0.1406^{+0.2502}_{-0.1179}$ & $4.509^{+0.03306}_{-0.008636}$ & $1.480^{+0.01881}_{-0.05027}$ & $4.708^{+0.02907}_{-0.008066}$ \\
XTE & $0.004583^{+0.01141}_{-0.003876}$ & $0.4773^{+0.5964}_{-0.4450}$ & $8.724^{+3.704}_{-3.538}$ & $0.5701^{+0.7706}_{-0.5324}$ & $5.371^{+1.227}_{-1.175}$ \\
GRS & $0.002276^{+0.005671}_{-0.001875}$ & $0.2875^{+0.2420}_{-0.2383}$ & $12.37^{+2.413}_{-2.239}$ & $0.2768^{+0.3900}_{-0.2560}$ & $4.031^{+0.6212}_{-0.6265}$ \\
H1743 & $0.002328^{+0.005713}_{-0.001926}$ & $0.2276^{+0.1391}_{-0.1328}$ & $11.05^{+2.264}_{-2.321}$ & $0.2775^{+0.3893}_{-0.2569}$ & $4.028^{+0.6051}_{-0.6127}$ \\
M87 & $0.0009268^{+0.002334}_{-0.0007832}$ & $0.2442^{+0.1505}_{-0.1344}$ & $419.4^{+21.94}_{-23.76}$ & $0.2165^{+0.3296}_{-0.2026}$ & $3.843^{+0.4917}_{-0.4873}$ \\
\hline
\multicolumn{6}{|c|}{\textbf{ER4 Model}} \\
\hline
Sgr A* & $0.0061^{+0.005767}_{-0.004402}$ & $0.6108^{+0.3631}_{-0.2804}$ & $3.938^{+0.04562}_{-0.06319}\times10^6$ & $0.8879^{+0.5817}_{-0.7419}$ & $5.787^{+0.7547}_{-1.093}$ \\
GRO & $0.005233^{+0.001028}_{-0.002036}$ & $0.1531^{+0.1607}_{-0.1381}$ & $4.511^{+0.03673}_{-0.009663}$ & $1.478^{+0.02165}_{-0.05633}$ & $4.708^{+0.0386}_{-0.007599}$ \\
XTE & $0.008909^{+0.007086}_{-0.008145}$ & $0.6038^{+0.6133}_{-0.5714}$ & $10.07^{+2.570}_{-4.884}$ & $0.4937^{+0.5825}_{-0.4559}$ & $5.322^{+1.276}_{-1.126}$ \\
GRS & $0.004396^{+0.003552}_{-0.003933}$ & $0.2973^{+0.3039}_{-0.2482}$ & $13.27^{+1.642}_{-3.138}$ & $0.2398^{+0.2984}_{-0.2190}$ & $3.977^{+0.6755}_{-0.5722}$ \\
H1743 & $0.004498^{+0.003543}_{-0.004035}$ & $0.2310^{+0.1788}_{-0.1361}$ & $11.90^{+1.542}_{-3.170}$ & $0.2398^{+0.2981}_{-0.2193}$ & $3.986^{+0.6476}_{-0.5702}$ \\
M87 & $0.001808^{+0.001453}_{-0.001657}$ & $0.2498^{+0.1880}_{-0.1400}$ & $427.3^{+15.28}_{-31.70}$ & $0.1857^{+0.2440}_{-0.1718}$ & $3.806^{+0.5290}_{-0.4500}$ \\
\hline
\end{tabular}
\label{tab:all-models}
\end{table*}

Table \ref{tab:all-models} summarizes the best-fit estimates and associated uncertainties of black hole parameters derived from five theoretical models which are RP, WD, ER2, ER3, and ER4 models within the framework of $f(R,T)$ gravity coupled to nonlinear electrodynamics (NED). The gravitational coupling parameter $\beta$, which characterizes the matter-geometry interaction in the modified gravity theory, shows moderate variations across different models. In the RP model, $\beta$ reaches comparatively larger values, ranging between $\sim 0.80$ and $1.88$, suggesting a stronger coupling between matter and curvature. The ER2 and ER3 models yield moderately constrained values, typically within $\sim 0.24$–$0.76$, with many values remaining below unity, indicating relatively weaker gravitational modifications. The ER4 model exhibits the most stringent bounds, with $\beta$ tightly confined to the range $\sim 0.15$–$0.61$, implying minimal deviations from GR. The WD model, in contrast, allows for the broadest span of $\beta$, ranging from $\sim 0.18$ to $0.74$, suggesting model flexibility and potential source-specific effects.
\\
Meanwhile, the nonlinearity parameter $\alpha$, which governs the strength of the NED correction, consistently remains small across all models and sources, indicating that the non-linear electromagnetic modifications are present but perturbative in nature. Analyzing the charge-to-mass ratio $Q/M$, some intriguing behaviors emerge, particularly in the GRO source under ER3 and ER4 models, where $Q/M$ exceeds unity. This hints at effective overcharged configurations shaped by the interplay of NED and modified gravity. The normalized orbital radius $r/M$ appears relatively stable, with only slight model-to-model variation, suggesting minor differences in the underlying effective potential.
\\
Overall, the analysis highlights the distinct and complementary roles of $\alpha$ and $\beta$: while $\alpha$ modulates the electromagnetic sector through nonlinear effects, $\beta$ encapsulates the strength of matter-curvature coupling in $f(R,T)$ gravity. Their combined influence governs the behavior of quasi-periodic oscillations (QPOs), offering valuable insights into the structure of black hole spacetimes in non-standard gravitational theories.\\
\section{Summary and Concluding Remarks}\label{secVI}

\begin{figure*}[t]
    \centering
    \subfloat[M82 X-1]{%
        \includegraphics[width=0.45\textwidth]{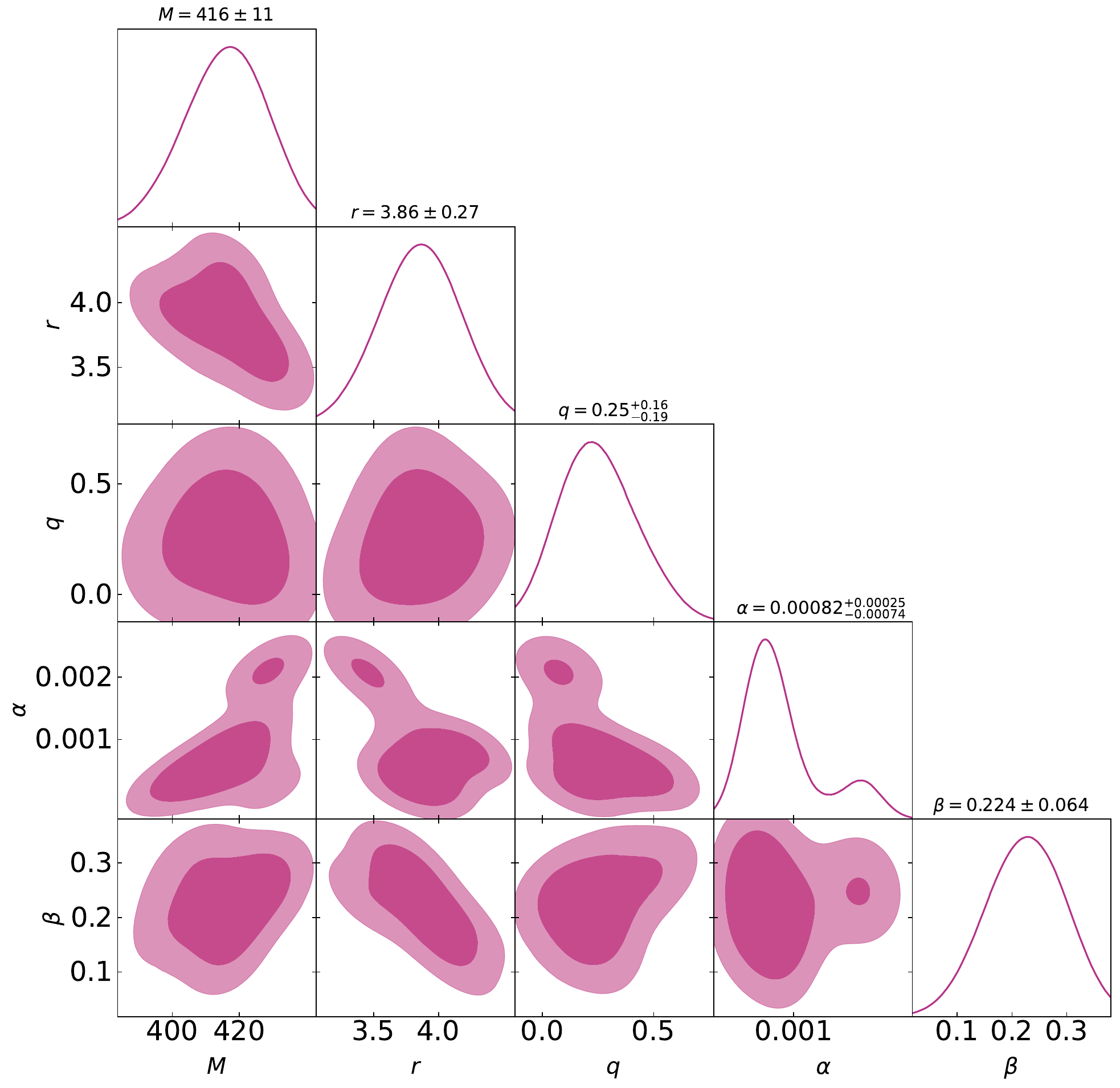}
    }\hfill
    \subfloat[Sgr A*]{%
       \includegraphics[width=0.45\textwidth]{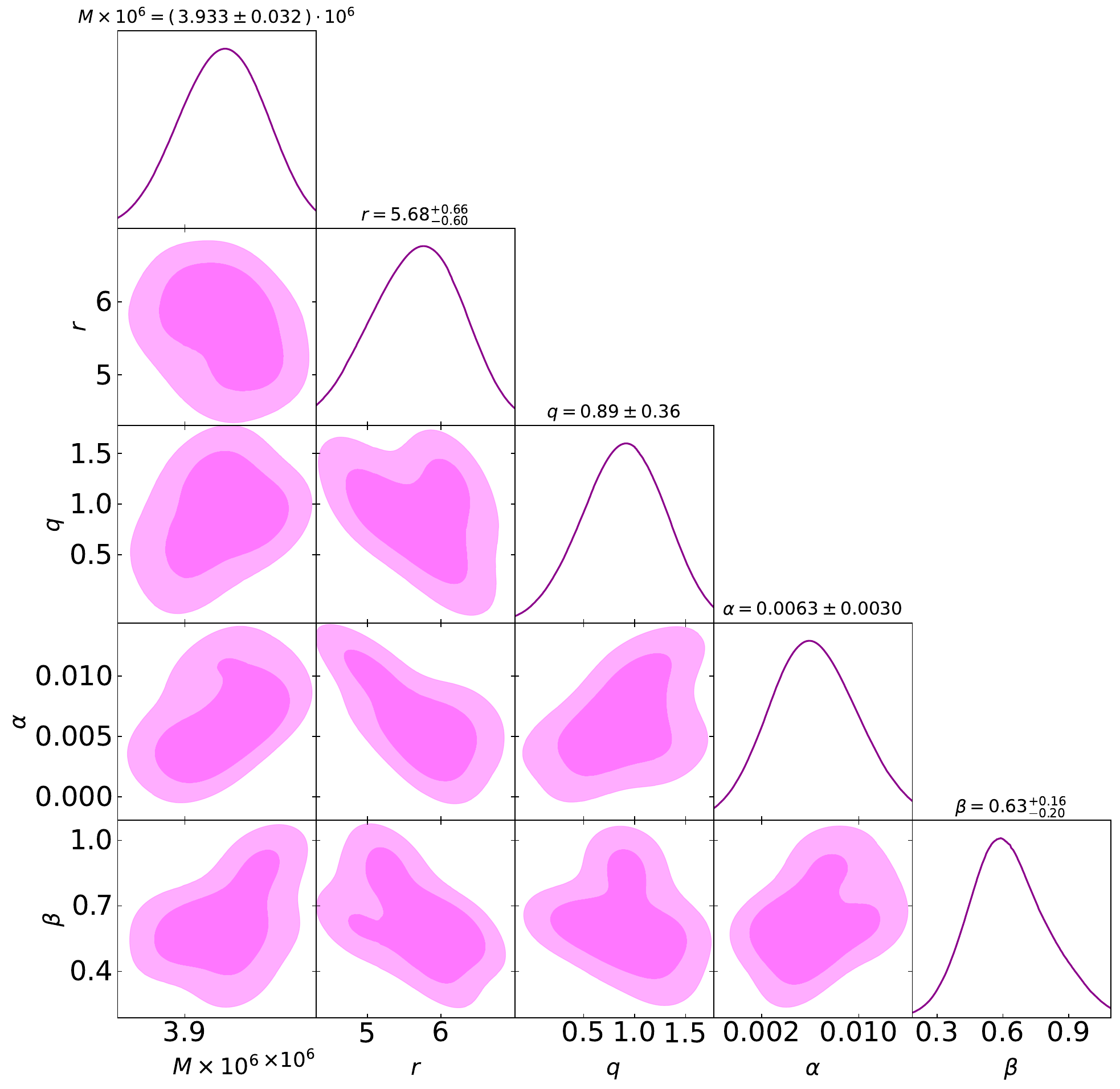}
    }
    \caption{Corner plots showing posterior distributions for the parameters \(M\), \(b\), \(Q/M\), and \(r/M\) obtained from the MCMC analysis for each black hole source.}
    \label{fig13}
\end{figure*}

In this study, we have explored the dynamics of circular motion and small oscillations of test particles in the gravitational field of a black hole solution governed by nonlinear electrodynamics (NED) within the framework of $f(R,T)$ gravity, with a particular emphasis on quasi-periodic oscillation (QPO) phenomena. Beginning with the derivation of the equations of motion, we examined the structure of the effective potential associated with circular geodesics. Our analysis reveals that, for a neutral test particle, an increase in the nonlinearity parameter $\alpha$ leads to an enhancement in the maximum of the effective potential relative to the Reissner–Nordstr\"om (RN) black hole. In contrast, increasing the coupling parameter $\beta$ results in a suppression of the peak, highlighting the competing influences of nonlinear electrodynamics and matter-curvature interactions on the orbital dynamics.\\
We also analyzed the behavior of specific energy and angular momentum for particles in circular orbits and observed that both quantities decrease as the nonlinear electrodynamics (NED) parameter $\alpha$ increases. This indicates that the orbits become more tightly bound in the presence of stronger nonlinear electromagnetic effects, causing the location of stable circular orbits to shift outward. Conversely, an increase in the gravitational coupling parameter $\beta$ leads to higher values of both specific energy and angular momentum, suggesting a weakening of the binding energy and a shift toward less tightly bound orbits. While a reduction in energy and angular momentum generally favors stability, excessive suppression due to strong nonlinearity can shrink the stable orbital region, potentially leading to dynamical instabilities. These deviations from the standard Reissner–Nordstr\"om (RN) case underscore the significant role that both NED corrections and matter-geometry coupling in $f(R,T)$ gravity play in shaping the dynamics of particles around black holes.
\\
We next investigated the evolution of the innermost stable circular orbit (ISCO) in the spacetime of a black hole governed by nonlinear electrodynamics (NED) within the framework of $f(R,T)$ gravity. Figure \ref{fig6} illustrates how the ISCO radius responds to variations in the gravitational coupling parameter $\beta$, across multiple values of the NED parameter $\alpha$. For all examined cases, the ISCO radius exhibits a clear monotonic increase with increasing $\beta$, indicating that stronger matter-curvature coupling systematically shifts the region of orbital stability outward. This trend persists consistently across different values of $\alpha$, highlighting a robust influence from the geometry-matter interaction intrinsic to modified gravity.
\\
Additionally, for fixed values of $\beta$, an increase in the NED parameter $\alpha$ also leads to a gradual outward displacement of the ISCO radius. This behavior reflects the impact of enhanced nonlinearity in the electromagnetic sector, which alters the underlying spacetime geometry and extends the stable orbital region farther from the black hole.
\\
%Overall, our findings reveal that both the gravitational coupling $\beta$ and the NED parameter $\alpha$ play crucial and systematic roles in shaping the dynamics of circular orbits. These results emphasize how the combined effects of nonlinear electromagnetic fields and matter-curvature interactions in $f(R,T)$ gravity modify the properties of black hole spacetimes, producing measurable deviations from standard Reissner–Nordström solutions.

We further extended our analysis to investigate quasi-periodic oscillations (QPOs) by examining epicyclic motion within several theoretical frameworks namely, the  Relativistic Precession (RP), Warped Disk (WD), and Epicyclic Resonance (ER2–ER4) models. Particular attention was given to resonance conditions corresponding to characteristic frequency ratios of $3\!:\!2$, $4\!:\!3$, and $5\!:\!4$, which are commonly observed in twin-peak QPOs.
\\
Figure~\ref{fig11} depicts the dependence of the QPO-generating orbital radius (normalized by $r/M$) on the gravitational coupling parameter $\beta$ for each model. The figure reveals that the WD model consistently yields resonance orbits at the largest radial distances, indicating that QPOs form farther from the black hole in this configuration compared to the others. The RP model follows closely, producing the next largest QPO-generating radii. In contrast, the ER2 model results in the most compact resonance radii among all five models, indicating that the corresponding oscillations are confined closer to the black hole horizon.
\\
Interestingly, the behavior of the resonance radius with respect to $\beta$ is non-monotonic across all models, suggesting that the coupling between matter and curvature in the $f(R,T)$ framework introduces complex modifications to the underlying spacetime structure. This complexity affects the location of resonance orbits in a nontrivial way, underlining the sensitivity of QPO phenomena to both gravitational and electromagnetic sector modifications. 
\\
To conclude our investigation, we carried out a comprehensive Markov Chain Monte Carlo (MCMC) analysis to constrain the parameter set $\{M, \beta, \alpha, Q/M, r/M\}$ characterizing the NED black hole spacetime in $f(R,T)$ gravity, using observational quasi-periodic oscillation (QPO) data from six prominent astrophysical black hole sources. These sources encompass a broad spectrum of mass scales, including stellar-mass black holes—GRO J1655–40, XTE J1550–564, GRS 1915+105, and H 1743–322—the intermediate-mass candidate M82 X-1, and the supermassive black hole at the Galactic center, Sgr A*.
\\
For the purpose of parameter estimation, we employed the Relativistic Precession (RP) model under the 3:2 resonance condition as a representative framework to fit the observed QPO frequencies. Additionally, analogous fits were performed for the remaining five theoretical models, and all their outcomes are systematically compiled in the accompanying Table \ref{tab:all-models}. These findings provide a comparative perspective on how different modeling frameworks within nonlinear electrodynamics and modified $f(R,T)$ gravity influence the inferred black hole parameters across various mass regimes.
\\
As a final and significant observation, we underscore that the NED black hole solution examined in this study exhibits the correct asymptotic behavior under physically meaningful limits. Notably, in the absence of electric charge, the metric consistently reduces to the well-established Schwarzschild geometry, confirming its compatibility with neutral black hole configurations. Moreover, in the simultaneous limit $\alpha \to 0$ and $\beta \to 1/2$, the solution naturally converges to the classical Reissner–Nordstr\"om spacetime. This seamless recovery of known solutions reinforces the theoretical robustness of the model and highlights its consistency within the broader framework of general relativity and nonlinear electrodynamics in modified gravity.
 The dynamical behavior of the system reveals several notable features that emerge depending on the values of the parameters $\alpha$ and $\beta$. Starting with the structure of the effective potential and extending through the profiles of specific energy and angular momentum, we find that increasing both $\alpha$ and $\beta$ progressively drives the black hole dynamics away from the standard Reissner–Nordstr\"om (RN)-like behavior. In contrast, for smaller values of the NED parameter $\alpha$ and gravitational coupling parameter $\beta$, the spacetime retains features more closely resembling those of the RN solution. This transition highlights the sensitivity of particle dynamics to nonlinear electrodynamic corrections and matter-curvature coupling in the $f(R,T)$ framework. 
This trend is clearly manifested in our investigation of the innermost stable circular orbit (ISCO) within the framework of nonlinear electrodynamics (NED) in $f(R,T)$ gravity. Our results show that the ISCO radius increases with the gravitational coupling parameter $\beta$, and this outward shift becomes more pronounced for higher values of the NED parameter $\alpha$. This indicates a synergistic influence of both geometry-matter coupling and nonlinear electromagnetic effects on the orbital stability region, pushing the ISCO farther from the black hole as these parameters increase.
In contrast, our analysis of QPO-generating orbits reveals that the resonance radius remains largely insensitive to variations in $\beta$, with no significant shift observed across the models. This distinction underscores the differing sensitivities of ISCO and QPO observables to the underlying spacetime structure.
To quantitatively assess these behaviors, we performed a Markov Chain Monte Carlo (MCMC) analysis using QPO data from black hole systems across the stellar, intermediate, and supermassive mass regimes. The inferred values of the NED parameter $\alpha$ were consistently found to be very small—typically much less than unity—indicating weak nonlinear electromagnetic corrections in the viable parameter space. Meanwhile, the gravitational coupling $\beta$ remained strictly below unity across all sources and models, suggesting modest but non-negligible modifications to the matter-curvature interaction. These findings collectively demonstrate that while ISCO properties are notably influenced by both $\alpha$ and $\beta$, the observable QPO structure is primarily governed by other aspects of the spacetime geometry and resonance dynamics.\\

%In conclusion, our study clearly demonstrates that the QPO characteristics of the NED black hole interpolate between the RN and Schwarzschild profiles, with the parameter \( b \) playing a pivotal role in governing this transition. Notably, both extremal limits, \( b \rightarrow 0 \) and \( b \rightarrow \infty \), produce well-behaved and physically meaningful profiles,. The most important outcome of our analysis is the clear signature of nonlinear electrodynamics (NED) on the QPO behavior of charged black holes. 
%This signature is reflected both in the theoretical aspects such as the effective potential, ISCO, and Keplerian frequency and in the observational context through the MCMC analysis based on QPO data. The constraints obtained on the NED parameter \( b \) remain consistently of order unity across different black hole sources, suggesting that the NED-induced modifications leave a subtle yet discernible signature on the QPO characteristics. These findings may indicate  the relevance of NED corrections in the context of black hole. \\
It is worth emphasizing that, although the observational QPO data utilized in this study originate from astrophysical sources believed to host rotating black holes, we have intentionally excluded the spin parameter from our theoretical modeling. This deliberate omission was aimed at isolating and assessing the intrinsic effects of nonlinear electrodynamics (NED) and matter-curvature coupling within the $f(R,T)$ gravity framework on the observed QPO frequencies. Our primary objective was to explore whether the inclusion of the NED nonlinearity parameter $\alpha$, alongside the gravitational coupling parameter $\beta$, could account for characteristic QPO features that are traditionally ascribed to black hole spin.
\\
Within this context, our analysis suggests that moderate values of the NED charge, modulated by small but non-zero $\alpha$, in conjunction with sub-unity values of $\beta$, may effectively emulate the frequency shifts typically attributed to rotational effects. This raises the intriguing possibility that spin-independent mechanisms such as nonlinear electromagnetic corrections and nonminimal matter-geometry interactions could mimic key observational signatures of rotating spacetimes. Such an interpretation provides a compelling alternative avenue for understanding QPO phenomena, particularly in scenarios where spin measurements are uncertain or ambiguous.
\\
It is important to acknowledge that the constraints on the nonlinear electrodynamics (NED) parameter $\alpha$ are likely to be affected by the inclusion of black hole spin in the theoretical framework. This interdependence is exemplified in Ref.~\cite{c2}, where the coupled influence of the NED parameter $k$ and the spin parameter $a$ on QPO modeling is systematically investigated. The findings of that study demonstrate a mutual sensitivity between these parameters, underscoring the nontrivial interplay between rotational dynamics and nonlinear electromagnetic corrections.
\\
In our current work, however, we have focused on the non-rotating case to isolate the effects of $\alpha$ and the gravitational coupling parameter $\beta$ within the context of $f(R,T)$ gravity. Incorporating rotation into this framework would necessitate a significant extension of the background spacetime to include a magnetically charged, rotating black hole solution consistent with both NED and modified gravity. Such a development is technically intricate and remains an open challenge. We recognize the importance of addressing this limitation and intend to pursue a detailed analysis of spin effects and their interaction with $\alpha$ and $\beta$ in future work.

\section*{Acknowledgements} 
B.H. gratefully acknowledges the financial support provided by the DST-INSPIRE Fellowship [IF220255], awarded by the Department of Science and Technology, Ministry of Science and Technology, Government of India.

\begin{widetext}
\section{Appendix}\label{A}
\begin{multline}\nonumber
\Omega_r^2 = \frac{1}{50 r^{14}} \left[ -300 M^2 r^{10} + 50 M r^{11} + 450 M Q^2 r^9 - 200 Q^4 r^8 
- 120 \alpha \beta Q^4 r^6 + 60 \alpha Q^4 r^6 + 120 \alpha \beta Q^6 r^4 - 60 \alpha Q^6 r^4 \right.\\
\left. \quad -10 \alpha \beta M Q^4 r^5 + 5 \alpha M Q^4 r^5 - 48 \alpha^2 \beta^2 Q^8 
+ 48 \alpha^2 \beta Q^8 - 12 \alpha^2 Q^8 \right]
\end{multline}
\end{widetext}

\end{document}